\documentclass[%
 reprint,
superscriptaddress,
 amsmath,amssymb,
 aps,
prstab,
]{revtex4-2}

\usepackage{graphicx}
\usepackage{subfigure}
\usepackage{dcolumn}
\usepackage{bm}
\usepackage{multirow,ragged2e}

\usepackage{color} 

\begin{document}

\preprint{APS/123-QED}

\title{Simulations and experimental results on beam-beam effects in SuperKEKB}

\author{Demin Zhou}
\email{dmzhou@post.kek.jp}
\affiliation{%
 KEK, 1-1 Oho, Tsukuba 305-0801, Japan 
}%
\affiliation{
 The Graduate University for Advanced Studies, SOKENDAI
}%
\author{Kazuhito Ohmi}%
\affiliation{%
 KEK, 1-1 Oho, Tsukuba 305-0801, Japan 
}%
\author{Yoshihiro Funakoshi}%
\affiliation{%
 KEK, 1-1 Oho, Tsukuba 305-0801, Japan 
}%
\author{Yukiyoshi Ohnishi}%
\affiliation{%
 KEK, 1-1 Oho, Tsukuba 305-0801, Japan 
}%
\affiliation{
 The Graduate University for Advanced Studies, SOKENDAI
}%


\author{Yuan Zhang}
\affiliation{
 Key Laboratory of Particle Acceleration Physics and Technology, Institute of High Energy Physics, \\
Chinese Academy of Sciences, 19(B) Yuquan Road, Beijing 100049, China
}%
\affiliation{
 University of Chinese Academy of Science, Beijing 100049, China
}%


\date{\today}

\begin{abstract}
The beam-beam interaction is one of the most critical factors determining the luminosity performance of colliders. As a circular collider utilizing the crab-waist scheme, multiple factors, such as beam-beam, crab waist, impedances, etc., interact to determine the luminosity of SuperKEKB. The interplay of these factors makes it challenging to predict luminosity via simulations. This paper presents recent advances in understanding the luminosity performance of SuperKEKB from beam-beam simulations and experiments. The key aspects affecting the luminosity of SuperKEKB, as well as the areas where further research is needed, are highlighted.
\end{abstract}

\maketitle


\section{\label{sec:Intro}Introduction}

The so-called ``nano-beam scheme'' was utilized in the design of the  SuperKEKB B-factory~\cite{Ohnishi2013PTEP, SuperKEKBTDR}. The collision scheme is similar to the crab waist (CW) scheme, which P. Raimondi originally proposed for SuperB~\cite{Raimondi2006, Raimondi2007arXiv, SuperBCDR}. The main difference is that the crab waist was not adopted as the baseline at SuperKEKB because it significantly reduces the dynamic aperture and lifetime~\cite{Morita:IPAC14-THPRI006} in the presence of realistic magnetic fields in the interaction region (IR)~\cite{Morita:IPAC11-THPZ006}. Even without the crab waist, it was found that there is a strong interplay between beam-beam interactions and lattice nonlinearity, causing a large luminosity degradation with the final design configurations (i.e., the vertical beta functions at the interaction point (IP) $\beta_y^*=0.27/0.3$ mm for the 4/7 GeV rings, respectively) of SuperKEKB~\cite{Zhou2015IPAC, ZhouICFABD2015}. The dominant sources of lattice nonlinearity in SuperKEKB were identified in the complicated IR with intentional orbit offsets in the final-focus superconducting magnets (the so-called QCS magnets)~\cite{Hirosawa:IPAC18-THPAK099}. Using a design lattice with $\beta_{y\pm}^*=1.08/1.2$ mm, simulations did not show such a large luminosity loss from this interplay~\cite{Zhou2015IPAC}.

SuperKEKB commissioning had three phases: Phase-1~\cite{Ohnishi2016, Funakoshi2016} (February - June 2016, without installation of the final focusing superconducting QCS magnets and roll-in of Belle II detector), Phase-2~\cite{Ohnishi2018} (February - July 2018, with QCS and Belle II, but without the vertex detector), and Phase-3~
\cite{Morita2019} (from March 2019 until present with the full Belle II detector). Beam commissioning without collisions in Phase-1 achieved small vertical emittances of less than 10 pm for both beams, which is essential for high luminosity. Machine tuning with collisions in Phase-2 confirmed the nano-beam collision scheme, i.e., collision with a large crossing angle and vertical beta function $\beta_y^*$ at the IP much smaller than the bunch length $\sigma_z$. Phase-3 commissioning started without the crab waist. In April 2020, the compact crab waist scheme, invented by K. Oide~\cite{Oide2016}, was successfully installed on SuperKEKB~\cite{Ohnishi2021EPJP}.

In Phase-2 commissioning without the crab waist, it was found that linear x-y coupling and dispersion at IP can severely degrade luminosity~\cite{Ohmi2018eeFACT}. The source of linear coupling was traced to unwanted skew-quadrupole components in the final focusing superconducting magnets. It was suspected that nonlinear chromatic and betatron couplings would be the next sources to explain the luminosity degradation. However, it was also suggested that nonlinear optical aberrations at the IP might be extremely large, which was inconsistent with optics measurements ~\cite{Ohmi2018eeFACT}. The coherent beam-beam head-tail instability (BBHTI, also called coherent X-Z instability in the literature) ~\cite{Ohmi2017PRL, Kuroo2018PRAB}, which cannot be suppressed by the crab waist, is potentially harmful to the luminosity performance. However, the BBHTI was observed in early Phase-2~\cite{Ohmi2018BBHTIeeFACT} but was suppressed in Phase-3 by squeezing $\beta_x^*$ plus careful optics tunings. Strong-strong beam-beam simulations showed that the beam-beam-driven betatron resonances were the most likely sources of luminosity degradation without the crab waist.

The uncontrollable blowup in vertical emittances severely limited the luminosity performance and motivated the installation of the crab waist to SuperKEKB.
Beam commissioning with the crab waist at SuperKEKB has been successful with $\beta_y^*=1$ and 0.8 mm~\cite{Ohnishi2021EPJP}. Experiments have shown that the crab waist effectively suppresses vertical blowup and allows larger beam currents to be stored in the rings~\cite{Funakoshi2022IPAC}, though luminosity performance is still worse than the predictions of simulations~\cite{Zhou:IPAC22-WEPOPT064}. On Jun. 22, 2022, a luminosity record of $4.71\times 10^{34}\text{ cm}^{-2}\text{s}^{-1}$ was achieved at SuperKEKB with $\beta_y^*=1$ mm and total beam currents $I_+/I_-=1.363/1.118$ A~\cite{Ohnishi_eeFACT2022}.

The orbit excursions in the IR magnets at full crossing angle essentially impact the linear optics. A correction scheme has been used in the design of IR optics to reduce the additional dispersion from the orbit excursions in the QCS magnets~\cite{Ohnishi2013PTEP, SuperKEKBTDR}. The beam-beam can interplay with the aberrations of the linear optics at the IP and cause luminosity degradation as investigated in Ref.~\cite{Ohmi2018eeFACT}. Special attention has been paid to optics tuning in machine operation, especially in the IR~\cite{sugimoto:eefact2022-tuxat0103}. In this paper, we assume the linear aberrations of IR optics are well understood and corrected.

This paper mainly addresses the beam-beam effects on achieving high luminosity with $\beta_y^*\geq 1$ mm at SuperKEKB. The paper is organized as follows. In Sec.~\ref{sec:Lum_formulations}, we present a compact overview of the formulae of luminosity and beam-beam tune shifts for flat-beam asymmetric colliders. The formulae in this section form the basis of the discussions on luminosity and beam-beam effects in SuperKEKB. The recent status of numerical codes and their applications to the beam-beam simulations for SuperKEKB is reviewed in Sec.~\ref{sec:Status_BB_simulation}. The experimental measurements of luminosity and their comparisons with simulations are documented in Sec.~\ref{sec:Lum_Performance}. The sources of luminosity degradation in machine operation are the main focus of Sec.~\ref{sec:Source_of_Lum_Degradation}. Finally, we summarize our findings in Sec.~\ref{sec:Summary} and also give an outlook of future directions to achieve higher luminosity performance in SuperKEKB.

\section{\label{sec:Lum_formulations}Formulae of luminosity and beam-beam tune shifts}
The luminosity of a collider can be calculated by performing the overlap integral of the 3D distributions of the colliding beams~\cite{HerrCAS2006}
\begin{equation}
    L=N_+N_-f_c K \int d^3\Vec{x} ds_0
    \rho_+(\Vec{x},-s_0) \rho_-(\Vec{x},s_0),
\end{equation}
with $f_c$ the collision frequency, $N_\pm$ the bunch populations, $\rho_\pm(\Vec{x},\pm s_0)$ the spatial distribution of the beams, and $K=\sqrt{(\Vec{v}_+-\Vec{v}_-)^2-(\Vec{v}_+\times\Vec{v}_-)^2/c^2}$ the kinematic factor. For SuperKEKB, the kinematic factor can be approximated by $K\approx 2c\cos^2\frac{\theta_c}{2}$, with $|\Vec{v}_\pm|=c$ and $\theta_c$ the full crossing angle. Gaussian distributions are often used to describe the beams
\begin{equation}
    \rho(x,y,s,s_0)=
    \frac{e^{-\frac{x^2}{2\sigma_{x}^2(s)}-\frac{y^2}{2\sigma_{y}^2(s,x)}-\frac{(s-s_0)^2}{2\sigma_{z}^2}}}{(2\pi)^{3/2}\sigma_{x}(s)\sigma_{y}(s,x)\sigma_{z}}
    \label{eq:GaussianChargeDensity}
\end{equation}
in the beams' frames. Here the transverse beam sizes $\sigma_{x,y}$ are written as functions of the longitudinal offset because of hourglass effects
\begin{equation}
    \sigma_x(s) = \sigma_x^* \sqrt{1+s^2/\beta_x^{*2}},
\end{equation}
\begin{equation}
    \sigma_y(s,x) = \sigma_{y}^*\sqrt{1+\left( s+R_{CW} x/\tan\theta_c \right)^2/\beta_y^{*2}},
\end{equation}
with $\beta_{x,y}^*$ the beta functions at the IP, $\sigma_{x,y}^*=\sqrt{\beta_{x,y}^*\epsilon_{x,y}}$ the beam sizes at IP. The parameter $R_{CW}$ is the crab waist ratio, with an arbitrary value of 1 for a full crab waist and 0 for no crab waist. The luminosity can be written as
\begin{equation}
    L=  \frac{N_bI_{b+}I_{b-} R_{HC}}{2\pi e^2f_0 \Sigma_x^* \Sigma_y^*} =L_0 R_{HC},
    \label{eq:Lum_RHC}
\end{equation}
where $\Sigma_u^*=\sqrt{\sigma_{u+}^{*2}+\sigma_{u-}^{*2}}$ with $u=x,y$, $R_{HC}$ the geometric factor representing the effects of crossing angle, hourglass effect, and the crab waist. For $\theta_c=0$, the explicit formula of $R_{HC}$ was given in Ref.~\cite{Furman:1991cr}. It can be extended to the case with finite crossing angle~\cite{Hirata1995PRL, peng2015luminosity} and craib waist~\cite{Dikansky2009NIMA}. The nominal luminosity $L_0$ is a function of the number of bunches $N_b$, the bunch currents $I_{b\pm}$, the transverse beam sizes at IP, and the revolution frequency $f_0$.

With $R_{CW}=0$ and flat-beam condition $\sigma_y^* \ll \sigma_x^*$, the geometric factor can be approximated by~\cite{Hirata1995PRL}
\begin{equation}
    R_{HC}\approx
    \sqrt{\frac{2}{\pi}} a e^b K_0(b),
    \label{eq:Hourglass_factor_with_crossing_angle}
\end{equation}
where $K_0(b)$ is the modified Bessel function of the second kind, which has the asymptotic property of $K_0(b)$$\approx$$e^{-b}\sqrt{\frac{\pi}{2b}}$ for large $b$. The parameters $a$ and $b$ are defined as
\begin{equation}
    a=
    \frac{\Sigma_y^*}{\Sigma_z\Sigma_\beta^*},
\end{equation}
\begin{equation}
    b=a^2 \left( 1+ \frac{\Sigma_z^2}{\Sigma_x^{*2}}\tan^2 \frac{\theta_c}{2} \right),
    \label{eq:b_parameter}
\end{equation}
with the quantities of $\Sigma_\beta^*$$=$$\sqrt{\sigma_{y+}^{*2}/\beta_{y+}^{*2}+\sigma_{y-}^{*2}/\beta_{y-}^{*2}}$ and $\Sigma_z$$=$$\sqrt{\sigma_{z+}^{2}+\sigma_{z-}^{2}}$. Though Eq.~(\ref{eq:Hourglass_factor_with_crossing_angle}) has the same form as in Ref.~\cite{Hirata1995PRL}, here the parameters $a$ and $b$ are extended to incorporate asymmetric beams.

From the geometric factor Eq.~(\ref{eq:Hourglass_factor_with_crossing_angle}), we can recognize three parameters that fundamentally define the luminosity and also the physics of beam-beam interaction in flat-beam asymmetric colliders:
\begin{itemize}
    \item $\Phi_{XZ}$=$\frac{\Sigma_z}{\Sigma_x^*}\tan\frac{\theta_c}{2}$, the ratio of dimensions of effective bunch length and the horizontal beam size projected to the longitudinal direction (i.e., $\Sigma_x^*/\tan\frac{\theta_c}{2}$ interpreted as the overlapping length of the colliding beams). This parameter essentially determines the overlapping area of the colliding beams~\cite{Raimondi2007arXiv}. For symmetric beams, it reduces to the well-known Piwinski angle $\Phi_P=\frac{\sigma_z}{\sigma_x^*}\tan\frac{\theta_c}{2}$.
    \item $\Phi_{HC}$=$\frac{\Sigma_y^*}{\Sigma_\beta^*\Sigma_x^*}\tan\frac{\theta_c}{2}$. When $\beta_{y+}^*$=$\beta_{y-}^*$, it reduces to $\Phi_{HC}$=$\frac{\beta_{y}^*}{\Sigma_{x}^*}\tan\frac{\theta_c}{2}$, the ratio of dimensions of vertical beta function at the IP and the overlapping length of the colliding beams. It can be taken as the hourglass factor for the crab-waist collision scheme.
    \item $\Phi_{H}$=$a$=$\frac{\Sigma_y^*}{\Sigma_z\Sigma_\beta^*}$. When $\beta_{y+}^*$=$\beta_{y-}^*$, it reduces to $\Phi_{H}$=$\beta_y^*/\Sigma_z$, the ratio of dimensions of vertical beta function at the IP and the effective bunch length. Note that, for symmetric beams, $\beta_y^*/\sigma_z$=$\sqrt{2}\Phi_{H}$ is the hourglass parameter, which defines the achievable $\beta_y^*$ in colliders with small Piwinski angles.
\end{itemize}
With the above definitions, the parameter $b$ of Eq.~(\ref{eq:b_parameter}) can be rewritten as
\begin{equation}
    b=\Phi_{H}^2 \left( 1+\Phi_{XZ}^2 \right)=\Phi_{H}^2 + \Phi_{HC}^2.
\end{equation}
With this formulation and the modified Bessel function $K_0(b)$, we can easily see how the luminosity of flat-beam colliders is related to the geometric parameters for different collision schemes:
\begin{itemize}
    \item For colliders with head-on collision or small Piwinski angle, there is $\Phi_{XZ}\ll 1$, and then $b\propto \Phi_H^2$. Consequently, there is $R_{HC}\approx 1$ when $b \gtrsim 1$ according to Eq.~(\ref{eq:Hourglass_factor_with_crossing_angle}). It implies that when $\Phi_H\gtrsim 1$, the hourglass effects on luminosity are negligible. This converts to the hourglass condition $\beta_y^*\gtrsim \sigma_z$ for colliders with small Piwinski angle.
    \item For colliders with large Piwinski angle (such as SuperB and SuperKEKB), there is $\Phi_{P} \gg 1$, and then $b \approx\Phi_{HC}^2$. Consequently, $R_{HC}$$\approx$ $\Phi_H/\Phi_{HC}$=$\Sigma_x^*/\left( \Sigma_z \tan \frac{\theta_c}{2} \right)$ when $\Phi_{HC} \gtrsim 1$, which is the condition of neglecting hourglass effects on luminosity for large crossing-angle collisions. It suggests that given horizontal beam sizes at the IP, $\beta_y^*$ needs to be larger than the overlapping length $\sigma_x^*/\tan\frac{\theta_c}{2}$. On the other hand, when $\beta_y^*$ is squeezed to achieve a certain target luminosity, the horizontal beam sizes at the IP must also be scaled down to avoid the unwanted hourglass effects. SuperKEKB was designed in the regime of $\Phi_P\gg 1$ and $\Phi_{HC}\lesssim 1$ (see Tab.~\ref{tb:parameters:LumTest}) where the hourglass effects on luminosity are not fully negligible. While in Phase-2 and Phase-3, SuperKEKB has been operating in the regime of $\Phi_P\gg 1$ and $\Phi_{HC}> 1$ (see Tabs.~\ref{tb:parameters:LumTest} and \ref{tb:parameters}), and the hourglass effects on luminosity are fairly negligible.
\end{itemize}
In general, for colliders with $b \gg 1$ (it can be realized by $\Phi_H\gg 1$ or by $\Phi_{HC}\gg 1$), the geometric factor reduces to
\begin{equation}
    R_{HC} \approx R_C =
    \frac{1}{\sqrt{1+ \frac{\Sigma_z^2}{\Sigma_x^{*2}}\tan^2 \frac{\theta_c}{2}}}.
    \label{eq:Cross_angle_factor}
\end{equation}
Consequently, the vertical beta function disappears, and the crossing angle alone determines the geometric factor. It suggests that the general condition for neglecting the vertical hourglass effect is $b\gtrsim 1$. When there is no hourglass effect in both $x$- and $y$-directions, there is exactly $R_{HC}=R_C$. Therefore, we can tentatively define the hourglass factor as
 \begin{equation}
     R_H=R_{HC}/R_C.
 \end{equation}

With full crab waist (i.e., $R_{CW}=1$) and a large Piwinski angle, the geometric factor $R_{HC}$ can be approximated as
\begin{equation}
    R_{HC}^{CW} \approx
    \frac{\Sigma_x^*\Sigma_z \tan \frac{\theta_c}{2}}{\Sigma_z^2\tan^2 \frac{\theta_c}{2}+
    \sigma_{x+}^*\sigma_{x-}^*} f(d),
    \label{eq:Hourglass_factor_with_crossing_angle_and_CW}
\end{equation}
with
\begin{equation}
    f(d)=\sqrt{\pi} d\cdot e^{d^2} \text{Erfc}(d),
\end{equation}
\begin{equation}
    d= \frac{\Sigma_y^*\Sigma_x^*}{\sqrt{2}\Sigma_\beta^* \sigma_{x+}^* \sigma_{x-}^*} \sin\theta_c.
\end{equation}
Here $\text{Erfc}(d)$ represents the complementary error function. For symmetric beams, $d$ reduces to $(\beta_y^*\sin\theta_c)/\sigma_x^*$, and Eq.~(\ref{eq:Hourglass_factor_with_crossing_angle_and_CW}) will have a simpler form, which can be derived from Eq.(15) of Ref.~\cite{Dikansky2009NIMA}.

It is seen that the horizontal beta function $\beta_x^*$ does not appear explicitly in Eqs. (\ref{eq:Hourglass_factor_with_crossing_angle}) and (\ref{eq:Hourglass_factor_with_crossing_angle_and_CW}), indicating that the horizontal hourglass effect can be neglected thanks to the flat-beam condition.

\begin{table}[hbt]
   \centering
   \caption{Machine parameters of SuperKEKB for tests of luminosity formulae. The set of ``Baseline design'' refers to Refs.~\cite{Ohnishi2013PTEP, SuperKEKBTDR} (Note that in Tab. 1 of Ref.~\cite{Ohnishi2013PTEP}, $\epsilon_y$=11.5 pm should be $\epsilon_y$=12.88 pm, according to Ref.~\cite{SuperKEKBTDR}.), and ``Phase-3'' refers to Tab. 2 (the column of 2021) of Ref.~\cite{Ohnishi2021EPJP}. The luminosity is calculated by Eqs. (\ref{eq:Lum_RHC}) and (\ref{eq:Hourglass_factor_with_crossing_angle}). The incoherent beam-beam tune shifts $\xi_{x,y}^i$ and $\xi_{x,y}^{ih}$  are calculated by Eq.~(\ref{eq:Incoherent_BB_tune_shift}) and by numerical integration of Eq.~(25) in Ref.~\cite{zhou2022formulae}, respectively.}
\begin{ruledtabular}
\begin{tabular}{ccccc}
\multirow{2}{*}{Parameters} &  \multicolumn{2}{c}{Baseline design} & \multicolumn{2}{c}{Phase-3 (2021)}\\
& LER & HER  & LER & HER\\
\hline
    $I_{b}$ (mA) & 1.44 & 1.04 & 0.673 & 0.585\\ 
    $\epsilon_x$ (nm) & 3.2 & 4.6 & 4.0 & 4.6\\ 
    $\epsilon_y$ (pm) & 8.64 & 12.88 & 52.5 & 52.5\\
    $\beta_x^*$ (mm) & 32 & 25 & 80 & 60\\
    $\beta_y^*$ (mm) & 0.27 & 0.3 & 1 & 1\\
    $\sigma_{z}$ (mm) & 6 & 5 & 4.6 & 5.1\\
    $N_b$              & \multicolumn{2}{c}{2500} &  \multicolumn{2}{c}{1174} \\
    $\xi_x^i$        & 0.0028 & 0.0012 & 0.0028 & 0.0030 \\
    $\xi_y^i$        & 0.078 & 0.074 & 0.0432 & 0.0314 \\
    $\xi_x^{ih}$        & 0.0019 & 0.0007 & 0.0028 & 0.0030 \\
    $\xi_y^{ih}$        & 0.088 & 0.078 & 0.0441 & 0.0318 \\
    $\Phi_{XZ}$              & \multicolumn{2}{c}{22.0} &  \multicolumn{2}{c}{11.6} \\
    $\Phi_{HC}$              & \multicolumn{2}{c}{0.8} &  \multicolumn{2}{c}{1.7} \\
    $L$ ($10^{34}\text{ cm}^{-2}\text{s}^{-1}$)    & \multicolumn{2}{c}{80.7} &  \multicolumn{2}{c}{3.0} \\
\end{tabular}
\end{ruledtabular}
\label{tb:parameters:LumTest}
\end{table}

The specific luminosity is defined as 
\begin{equation}
    L_{sp}=\frac{L}{N_bI_{b+}I_{b-}},
\end{equation}
which is a geometric parameter indicating the potential of a collider for generating collision events in particle detectors. Using the previous formulations, it can be expressed as
\begin{equation}
  L_{sp}= 
  \frac{L_0}{N_bI_{b+}I_{b-}} R_C R_H.
 \label{eq:Lsp}
\end{equation}
Considering a very large Piwinski angle $\Phi_P\gg 1$, the specific luminosity is approximated by
\begin{equation}
    L_{sp} \approx
    \frac{1}{2\pi e^2f_0 \Sigma_y^* \Sigma_z \tan \frac{\theta_c}{2}}.
    \label{eq:Lsp_Simple_approx1}
\end{equation}

For SuperKEKB, the validity of the aforementioned luminosity formulae was checked in Ref.~\cite{zhou2022formulae} by beam-beam simulations using the machine parameters of Tab.~\ref{tb:parameters:LumTest}. The main findings were 1) Without the crab waist, Eq.~(\ref{eq:Hourglass_factor_with_crossing_angle}) is a very good approximation of luminosity for the nano-beam scheme; 2) With the crab waist and a large Piwinski angle, Eq.~(\ref{eq:Hourglass_factor_with_crossing_angle_and_CW}) is a fairly good approximation to the luminosity for the crab waist scheme; 3) According to Eq.~(\ref{eq:Hourglass_factor_with_crossing_angle}), nonnegligible hourglass effect on luminosity appears and Eq.~(\ref{eq:Cross_angle_factor}) does not apply when $b < 1$ (i.e., the condition $b < 1$ results from squeezing $\beta_y^*$ or enlarging $\sigma_x^*$ until $K_0(b)$$\approx$$e^{-b}\sqrt{\frac{\pi}{2b}}$ is not valid; 4) The crab waist modifies the beam distribution, causing a luminosity gain of a few percent or less; 5) With operation conditions until June 2022 (i.e., $\beta_y^*\geq 1$ mm), the simple formula $L_0R_C$ is fairly good to describe the luminosity of SuperKEKB. Consequently, using this formula to estimate the beam sizes at the IP from measured luminosity is also valid.

 The beam-beam interaction will cause betatron tune shifts, which are important parameters for measuring the luminosity potential of a collider. The incoherent beam-beam tune shifts can be calculated from the beam-beam kick~\cite{Raimondi2003} on the on-axis particles. With the hourglass effect neglected and assumed Gaussian beams, they are given by
 \begin{equation}
    \xi_{u\pm}^i= \frac{r_e}{2\pi\gamma_\pm} \frac{N_\mp\beta_{u\pm}^*}{\overline{\sigma}_{u\mp}(\overline{\sigma}_{x\mp}+\overline{\sigma}_{y\mp})},
    \label{eq:Incoherent_BB_tune_shift}
\end{equation}
with $u=x,y$. In the case of a finite horizontal crossing angle, the beam sizes in the above equation are defined as $\overline{\sigma}_{y\pm}=\sigma_{y\pm}^*$, and $\overline{\sigma}_{x\pm}=\sqrt{\sigma_{z\pm}^2\tan^2\frac{\theta_c}{2}+\sigma_{x\pm}^{*2}}$. The formula is the same as that for a head-on collision, except that the horizontal beam size is modified. The incoherent beam-beam tune shifts depend on the opposite beam's bunch current and beam sizes.

With the hourglass effect taken into account, the beam-beam tune shift of on-axis particles (i.e., $\xi_{u\pm}^{ih}$) can be numerically calculated by integrating the $\beta$-function weighted beam-beam force along their path. For example, one can refer to Eq. (7) of Ref.~\cite{Valishev2013Fermi} to perform the numerical integration. We can define the hourglass factor for beam-beam tune shifts as
\begin{equation}
    R_{\xi u\pm}=\xi_{u\pm}^{ih}/\xi_{u\pm}^i.
\end{equation}
For SuperKEKB, the incoherent beam-beam tune shifts, i.e., $\xi_{x,y}^i$ by the simple estimate of Eq.~(\ref{eq:Incoherent_BB_tune_shift}) and $\xi_{x,y}^{ih}$ by numerical integration of Eq.~(25) in Ref.~\cite{zhou2022formulae}, are compared in Tabs.~\ref{tb:parameters:LumTest} and \ref{tb:parameters}. One can see that the two methods give results close to each other.

Empirically, we often calculate the vertical beam-beam parameter of flat beams from luminosity~\cite{Ohmi2004PRSTAB}
\begin{equation}
    L = \frac{1}{2e r_e} \frac{\gamma_\pm I_\pm}{\beta_{y\pm}^*} \xi_{y\pm}^L,
\end{equation}
with $I_\pm$ the total beam currents. Here the hourglass effects are resolved in the beam-beam parameter $\xi_{y\pm}^L$. In terms of incoherent beam-beam tune shifts, the luminosity Eq.~(\ref{eq:Lum_RHC}) can be expressed as
\begin{equation}
    L = \frac{1}{2e r_e} \frac{\gamma_\pm I_\pm}{\beta_{y\pm}^*} \xi_{y\pm}^i
      \frac{2\sigma_{y\mp}^*\left(\overline{\sigma}_{x\mp}+\sigma_{y\mp}^*\right)}{\Sigma_y^*\overline{\Sigma}_x} R_H,
\end{equation}
with $\overline{\Sigma}_x=\sqrt{\overline{\sigma}_{x+}^2+\overline{\sigma}_{x-}^2}$ the effective horizontal beam size. One can see that, for 3D Gaussian beams with identical sizes (i.e., $\sigma_{u+}^*=\sigma_{u-}^*$) and flat-beam condition (i.e, $\sigma_{y\pm}^*\ll \sigma_{x\pm}^*$), there is $\xi_{y\pm}^L=\xi_{y\pm}^i R_H$. When the hourglass effect is negligible, the relation is simple: $\xi_{y\pm}^L=\xi_{y\pm}^i$. Further correlation to the incoherent beam-beam tune shifts with the hourglass effect is
\begin{equation}
    \xi_{y\pm}^{ih} =\xi_{y\pm}^{i} R_{\xi y\pm}=\xi_{y\pm}^{L} R_{\xi y\pm}/R_H.
    \label{eq:BB_paramemter_with_Hourglass_effect1}
\end{equation}
Here $\xi_{y\pm}^{ih}$ is consistent with the definition of $\xi_y$ in Eq.(2.3) of Ref.~\cite{KEKB_Design_Report}, with the condition that the beam sizes of the two beams are equal.

Consider a very large Piwinski angle and assume that the hourglass effect is negligible; from Eq.~(\ref{eq:Incoherent_BB_tune_shift}), the incoherent vertical beam-beam tune shift can be simplified to
\begin{equation}
    \xi_{y\pm}^i \approx
    \frac{r_e}{2\pi ef_0 \gamma_\pm \tan \frac{\theta_c}{2} }
    \frac{I_{b\mp}\beta_{y\pm}^*}{\sigma_{y\mp}^* \sigma_{z\mp}}.
    \label{eq:Incoherent_BB_tune_shift_approx1}
\end{equation}
Furthermore, we assume that a balanced collision is achievable: $\beta_{y+}^*=\beta_{y-}^*=\beta_y^*$ and $\epsilon_{y+}=\epsilon_{y-}=\epsilon_y$. The above equation can then be rewritten as
\begin{equation}
    \xi_{y\pm}^i \approx
    \frac{r_e}{2\pi ef_0 \gamma_\pm \tan \frac{\theta_c}{2} }
    \frac{I_{b\mp}}{\sigma_{z\mp}}
    \sqrt{\frac{\beta_y^*}{\epsilon_y}}.
    \label{eq:Incoherent_BB_tune_shift_approx2}
\end{equation}
Suppose there is an upper limit on the achievable beam-beam tune shift (i.e., the beam-beam tune shift saturates to a certain value, and the collider reaches the so-called beam-beam limit). In that case, the above equation shows a constraint between the bunch current, the vertical emittance, and the vertical beta function at the IP. For example, to achieve the same beam-beam tune shift at a given bunch current, squeezing $\beta_y^*$ requires reducing the single-beam emittance $\epsilon_y$. On the other hand, if $\beta_y^*$ is fixed by optics design, the beam-beam limit suggests that an emittance blowup scaling by $\epsilon_y \propto I_{b\pm}^2$ is expected.

\begin{table}[hbt]
   \centering
   \caption{SuperKEKB machine parameters for $\beta_y^*$=2 mm on Jul. 1, 2019 and $\beta_y^*$=1 mm on Apr. 5, 2022, respectively. The luminosity is calculated by Eqs. (\ref{eq:Lum_RHC}) and (\ref{eq:Hourglass_factor_with_crossing_angle}). The incoherent beam-beam tune shifts $\xi_{x,y}^i$ and $\xi_{x,y}^{ih}$  are calculated by Eq.~(\ref{eq:Incoherent_BB_tune_shift}) and by numerical integration of Eq.~(25) in Ref.~\cite{zhou2022formulae}, respectively.}
\begin{ruledtabular}
\begin{tabular}{ccccc}
\multirow{2}{*}{Parameters} &  \multicolumn{2}{c}{2019.07.01} & \multicolumn{2}{c}{2022.04.05}\\
& LER & HER  & LER & HER\\
\hline
    $I_{b}$ (mA) & 0.51 & 0.51 & 0.71 & 0.57\\ 
    $\epsilon_x$ (nm) & 2.0 & 4.6 & 4.0 & 4.6\\ 
    $\epsilon_y$ (pm) & 40 & 40 & 30 & 35\\
    $\beta_x$ (mm) & 80 & 80 & 80 & 60\\
    $\beta_y$ (mm) & 2 & 2 & 1 & 1\\
    $\sigma_{z0}$ (mm) & 4.6 & 5.0 & 4.6 & 5.1\\
    $\nu_{x}$ & 44.542 & 45.53 & 44.524 & 45.532\\
    $\nu_{y}$ & 46.605 & 43.583 & 46.589 & 43.572\\
    $\nu_{s}$ & 0.023 & 0.027 & 0.023 & 0.027\\
    Crab waist ratio & 0 & 0 & 80\% & 40\% \\
    $N_b$              & \multicolumn{2}{c}{1174} &  \multicolumn{2}{c}{1174} \\
    $\xi_x^i$        & 0.0034 & 0.0023 & 0.0036 & 0.0024 \\
    $\xi_y^i$        & 0.0621 & 0.0386 & 0.052 & 0.044 \\
    $\xi_x^{ih}$        & 0.0034 & 0.0023 & 0.0036 & 0.0024 \\
    $\xi_y^{ih}$        & 0.0621 & 0.0383 & 0.0523 & 0.0446 \\
    $\Phi_{XZ}$              & \multicolumn{2}{c}{12.3} &  \multicolumn{2}{c}{11.7} \\
    $\Phi_{HC}$              & \multicolumn{2}{c}{3.6} &  \multicolumn{2}{c}{1.7} \\
    $L$ ($10^{34}\text{ cm}^{-2}\text{s}^{-1}$)    & \multicolumn{2}{c}{1.7} &  \multicolumn{2}{c}{3.9} \\
\end{tabular}
\end{ruledtabular}
\label{tb:parameters}
\end{table}

In Tab.~\ref{tb:parameters:LumTest}, it is shown that the hourglass effect modifies the vertical incoherent tune shifts $\xi_y$ by about 11\% and 5\% respectively for LER and HER (see the difference between $\xi_y^i$ and $\xi_y^{ih}$) with the baseline design configuration of SuperKEKB. Table ~\ref{tb:parameters} shows the typical machine parameters from the operation without the crab waist (2019.07.01) and with the crab waist (2022.04.05). One can see that, for the cases of $\beta_y^*\geq 1$ mm, the hourglass effect on the vertical incoherent tune shifts is negligible at SuperKEKB. Therefore we mainly refer to Eq.~(\ref{eq:Incoherent_BB_tune_shift}) for beam-beam tune shifts in the following discussions. The horizontal incoherent tune shifts are smaller than the vertical ones by one order and will not be discussed in detail in this paper.

\begin{figure}[htb]
   \centering
    \vspace{-1mm}
   \includegraphics*[width=80mm]{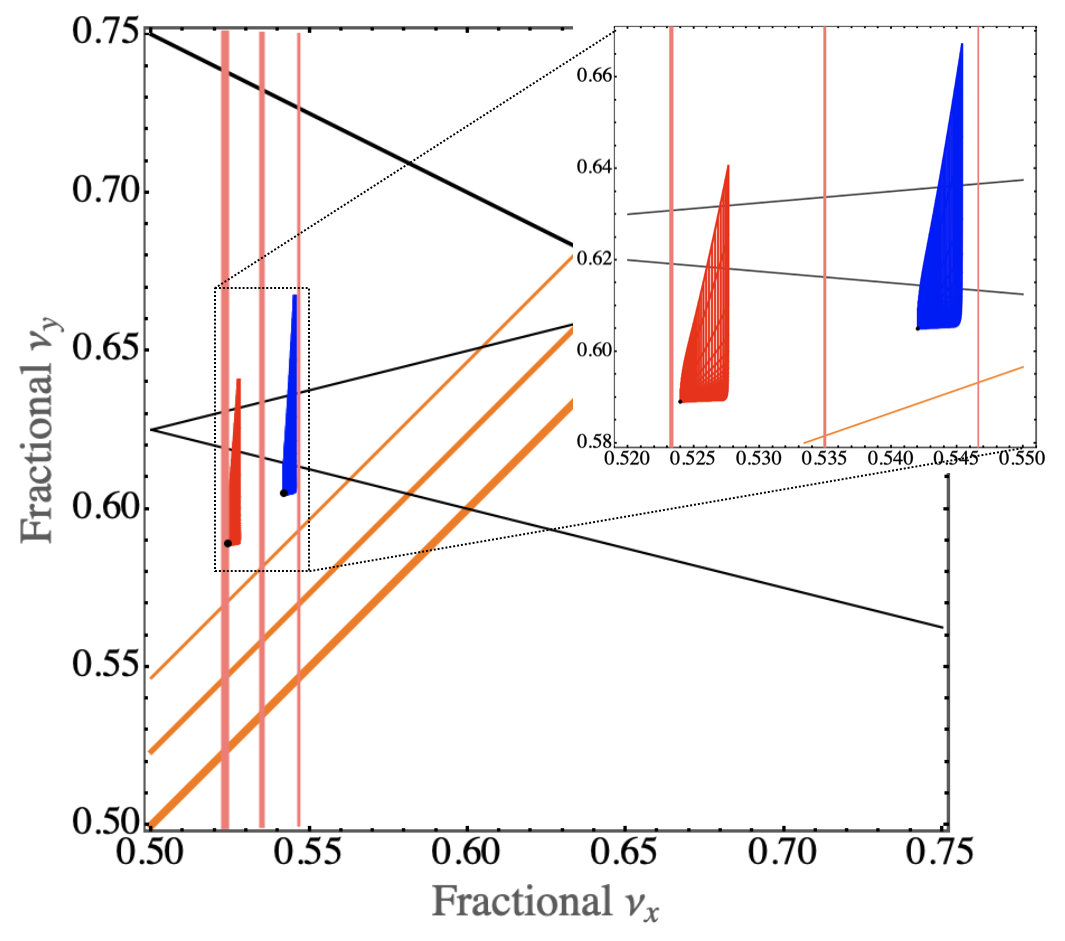}
    \vspace{-1mm}
   \caption{Beam-beam driven footprint of LER beam in the tune space with parameters of Tab.~\ref{tb:parameters}. The blue and red footprints represent 2019.07.01 and 2022.04.05, respectively. The black dots indicate the working points ($\nu_x,\nu_y$) shown in Tab.~\ref{tb:parameters}.}
   \label{fig:TuneFootprintLER}
    \vspace{-2mm}
\end{figure}

Using the parameters of Tab.~\ref{tb:parameters}, the beam-beam-induced footprints of the positron beam in the tune space are plotted in Fig.~\ref{fig:TuneFootprintLER} with solid lines indicating the important resonances. The linear and chromatic coupling resonances $\nu_x-\nu_y+k\nu_s=N$ are driven by machine imperfections. The resonances at $\nu_x\pm n\nu_y =N$ can be excited by the beam-beam interaction with a large crossing angle. The synchro-betatron resonances $2\nu_x-k\nu_s=N$ can be excited by both machine imperfections and the beam-beam interaction. Here the incoherent betatron and synchrotron tunes are used to describe the resonances. Transverse coupling impedances and beam-beam effects can cause shifts of the incoherent betatron tunes, and potential-well distortion from longitudinal impedance can cause a shift of the incoherent synchrotron tune. Therefore, as bunch currents change, the positions of relevant resonant lines also shift dynamically in the tune space. The rule of thumb is to find a working point to avoid obvious overlap between the beam's footprint and harmful resonances~\cite{Shatilov2017ICFABD}.

\section{\label{sec:Status_BB_simulation}Status of beam-beam simulations}

Beam-beam simulations for SuperKEKB have been intensively done since the design stage. Simulation codes include BBWS~\cite{ohmi:pac03-woaa005}, SAD~\cite{SADwebpage}, BBSS~\cite{Ohmi2000PRE,Ohmi2004PRL} and IBB~\cite{Zhang2005PRSTAB}. BBWS and BBSS were developed by K. Ohmi at KEK, and Y. Zhang developed IBB at IHEP. BBWS simulations use a weak-strong model for the beam-beam interaction, a one-turn matrix for lattice transformation, perturbation maps for linear and nonlinear machine imperfections, ideal crab waist, longitudinal and transverse beam coupling impedances, etc. SAD simulations use the weak-strong beam-beam model of BBWS and allow the loading of a full lattice, perturbation maps, etc. BBSS simulations use a strong-strong model for the beam-beam interaction and all features of BBWS. IBB is an MPI-based parallel strong-strong code and has similar features to BBSS.

SAD simulations are used to investigate the interplay of beam-beam and lattice nonlinearities~\cite{Zhou2015IPAC}. BBWS simulations have been frequently used to estimate luminosity performance and tune scans. BBSS and IBB simulations are essential in simulating the coherent beam-beam instabilities and have been used for investigating the interplay of beam-beam, impedances, and machine imperfections.

Currently, most strong-strong simulations using BBSS and IBB simulations for SuperKEKB utilize the so-called soft-Gaussian approach with beam-size parameters calculated from macro particles' coordinates. The bunches are sliced in the longitudinal direction, and the transverse sizes of each slice are calculated by fitting the transverse histogram or simply from RMS statistics. The number of macro particles is typically one million or more for each bunch to reduce statistical errors. Ideally, particle-in-cell (PIC) simulations are the most accurate among various models of beam-beam interaction. But for the case of SuperKEKB, PIC simulations are about 100 times slower than simulations using the soft-Gaussian approach.

\begin{figure}[htb]
   \centering
    \vspace{-2mm}
   \includegraphics*[width=70mm]{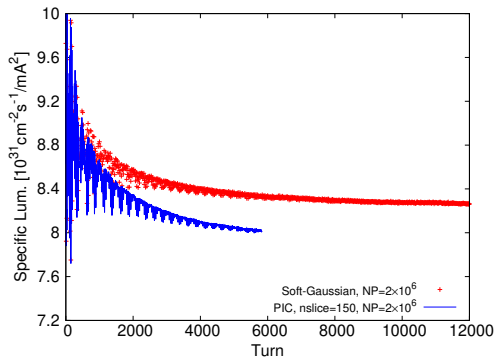}
    \vspace{-2mm}
   \caption{Simulated specific luminosity using BBSS code with the parameter set of 2022.04.05 in Tab.~\ref{tb:parameters} with the bunch currents and vertical emittances replaced by $I_{b+}/I_{b-}=1.0/0.8$ mA and $\epsilon_{y+}/\epsilon_{y-}$=20/35 pm.}
   \label{fig:Lsp_PIC_vs_Gaussian}
    \vspace{-4mm}
\end{figure}

Figure~\ref{fig:Lsp_PIC_vs_Gaussian} compares BBSS simulations using PIC and the soft-Gaussian approach. A PIC simulation took about 4 months for 6000 turns of tracking using 8 cores of a workstation with 3-GHz CPUs. On the other hand, a soft-Gaussian simulation for 12000 turns took about 40 hours using the same amount of CPU resources. It is noteworthy that the PIC simulation predicts a luminosity about 5\% lower than that by soft-Gaussian simulations. This difference is probably due to the fact that the crab waist causes deformation of the beam distribution, making a Gaussian approximation insufficient for the estimate of transverse beam sizes. Another issue is that numerical noise is always present in strong-strong simulations. The numerical noise affects the accuracy of the beam sizes in soft-Gaussian simulations and enhances the diffusion in the PIC simulations.

\section{\label{sec:Lum_Performance}Luminosity performance}

\subsection{\label{sec:Lum_Performance_wo_CW}Luminosity performance without crab waist}

From March 2018 to March 2020, SuperKEKB was operated with collisions but without a crab waist. During that time, many challenges were experienced:
1) Peak luminosity was much lower than predictions of beam-beam simulations~\cite{Ohmi2018eeFACT};
2) Severe vertical blowup with a threshold bunch current lower than 1 mA was observed even with the single-beam operation (no collision) in the LER~\cite{Terui2022IPAC}; 
3) The area with good luminosity in tune space was small compared with predictions of beam-beam simulation; 
4) The beam-induced backgrounds in Belle II were unexpectedly high~\cite{Shibata_IPAC2020}; 
5) The gain of luminosity via squeezing $\beta_{x,y}^*$ was small due to extra emittance blowup; 
6) It was difficult to operate the machine at the design working point (.53, .57) where beam-beam simulations predict the best luminosity performance (See Tab.~\ref{tb:parameters} for the working points without crab waist in July 2019.). Further information on SuperKEKB commissioning without the crab waist can be found in Refs.~\cite{Ohnishi2018, Morita2019, Shibata_IPAC2020}.

\begin{figure}[!htb]
   \centering
   \vspace{-9mm}
    \includegraphics*[width=80mm]{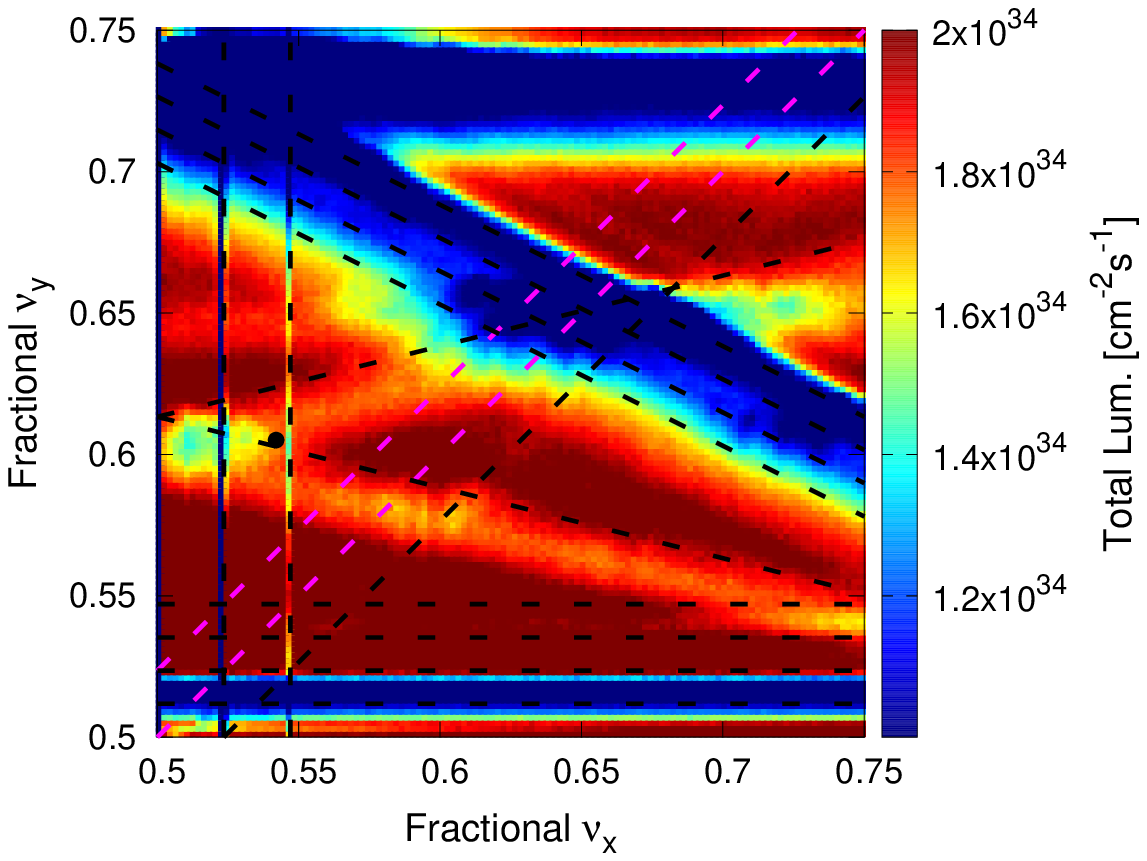}
    \vspace{0mm}
    \vspace{0mm}
    \includegraphics*[width=80mm]{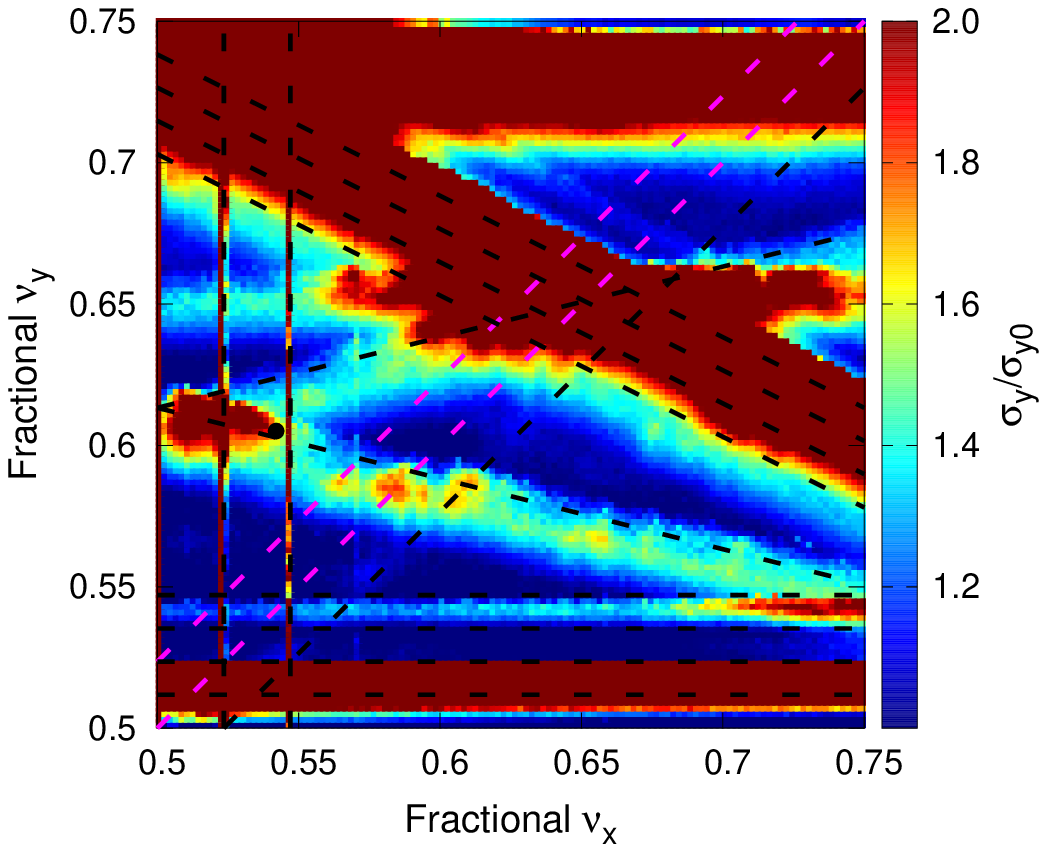}
    \vspace{-4mm}
   \caption{Tune scan of luminosity (upper) and vertical beam size (lower, normalized by $\sigma_{y0}$) for the parameter set of 2019.07.01 in Tab.~\ref{tb:parameters} with the LER as the weak beam in the BBWS simulation. Important resonant lines are plotted, and the black dot indicates the working point for machine operation.}
   \label{fig:Lum_tune_scan}
\end{figure}

\begin{figure}[htb]
   \centering
    \vspace{-2mm}
   \includegraphics*[width=70mm]{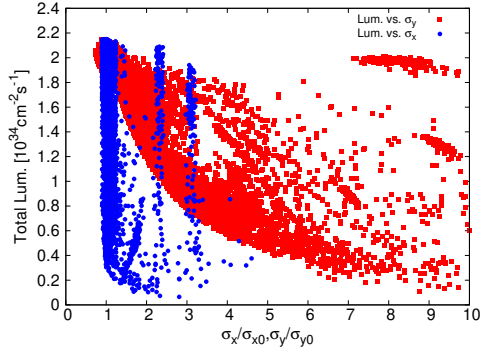}
    \vspace{-2mm}
   \caption{Correlation between the simulated total luminosity and horizontal/vertical beam sizes.}
   \label{fig:Lsp_vs_sigxy}
    \vspace{-4mm}
\end{figure}

Because of the large crossing angle, the beam-beam-driven resonances $\nu_x\pm 4\nu_y + \alpha =N$ (Here the parameter $\alpha$ scales as the vertical beam-beam tune shift.) have a strong impact on the vertical blowup observed at SuperKEKB without crab waist. This is illustrated by beam-beam simulations as shown in Fig.~\ref{fig:Lum_tune_scan} for a luminosity tune scan. The weak-strong BBWS code was used for the simulations with machine parameters referring to the LER set of 2019.07.01 in Tab.~\ref{tb:parameters}. The black dots represent the working point, which was optimized considering the overall performance (i.e., luminosity, background, injection efficiency, etc.). The same data of Fig.~\ref{fig:Lum_tune_scan} is plotted in Fig.~\ref{fig:Lsp_vs_sigxy} to show the relationship between the luminosity and horizontal/vertical beam sizes. It is seen that the luminosity strongly correlates with the vertical beam size. The 5th-order beam-beam resonances widen as beam currents increase, making it difficult to find a good working point to avoid beam-size blowups. The fractional vertical tune $\nu_y$ could not approach downward to the design value 0.57, partly affected by the chromatic coupling resonances $\nu_y-\nu_x-k\nu_s=N$ (magenta lines in the figures with $k=1,2$). The impedance effects also played a role in the choice of $\nu_y$, as will be addressed in detail in Sec.~\ref{sec:Lum_with_CW}.

\subsection{\label{sec:Lum_with_CW}Luminosity performance with crab waist}

\begin{figure}[htb]
   \centering
    \vspace{-2mm}
   \includegraphics*[width=70mm]{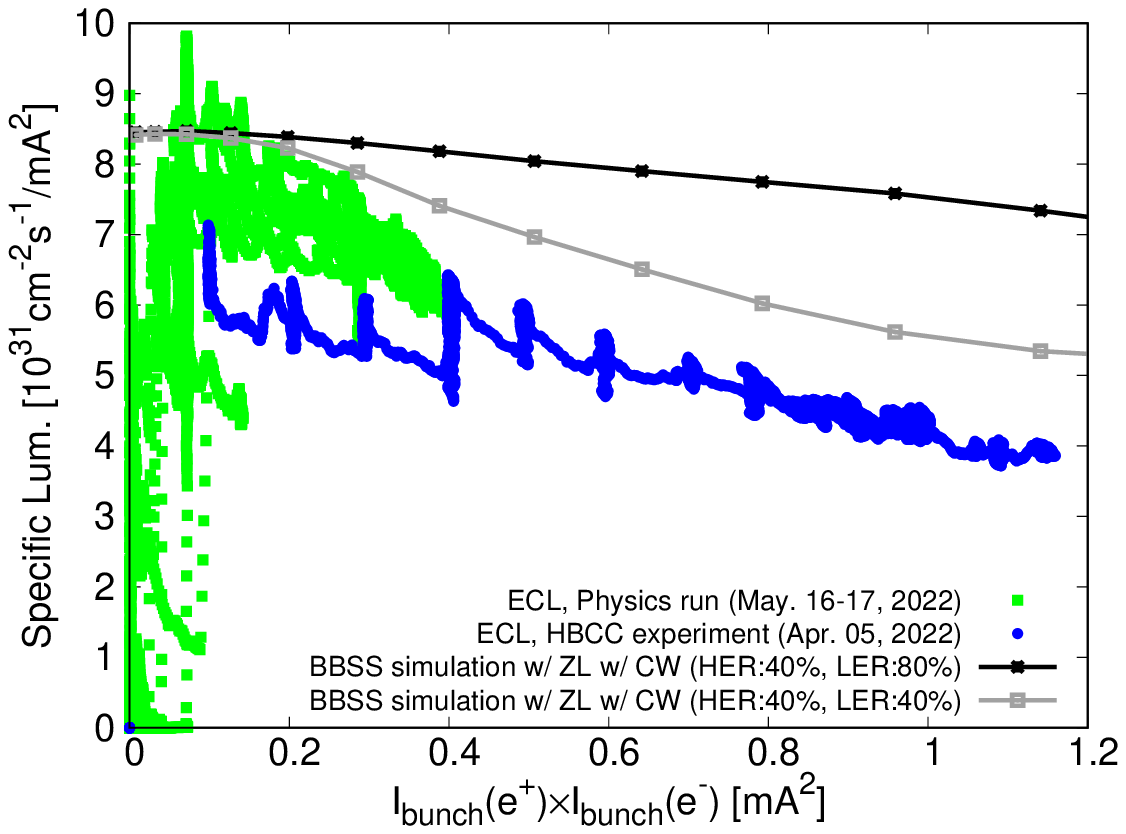}
    \vspace{0mm}
   \caption{Specific luminosity from HBCC machine study (blue dots) and physics run (green dots) measured by the ECL monitor in 2022, compared to predictions of BBSS simulations with the inclusion of longitudinal impedances.}
   \label{fig:Lum_Sim_vs_Exp_20220405}
\end{figure}

\begin{figure}[!htb]
   \centering
   \vspace{-5mm}
    \includegraphics*[width=70mm]{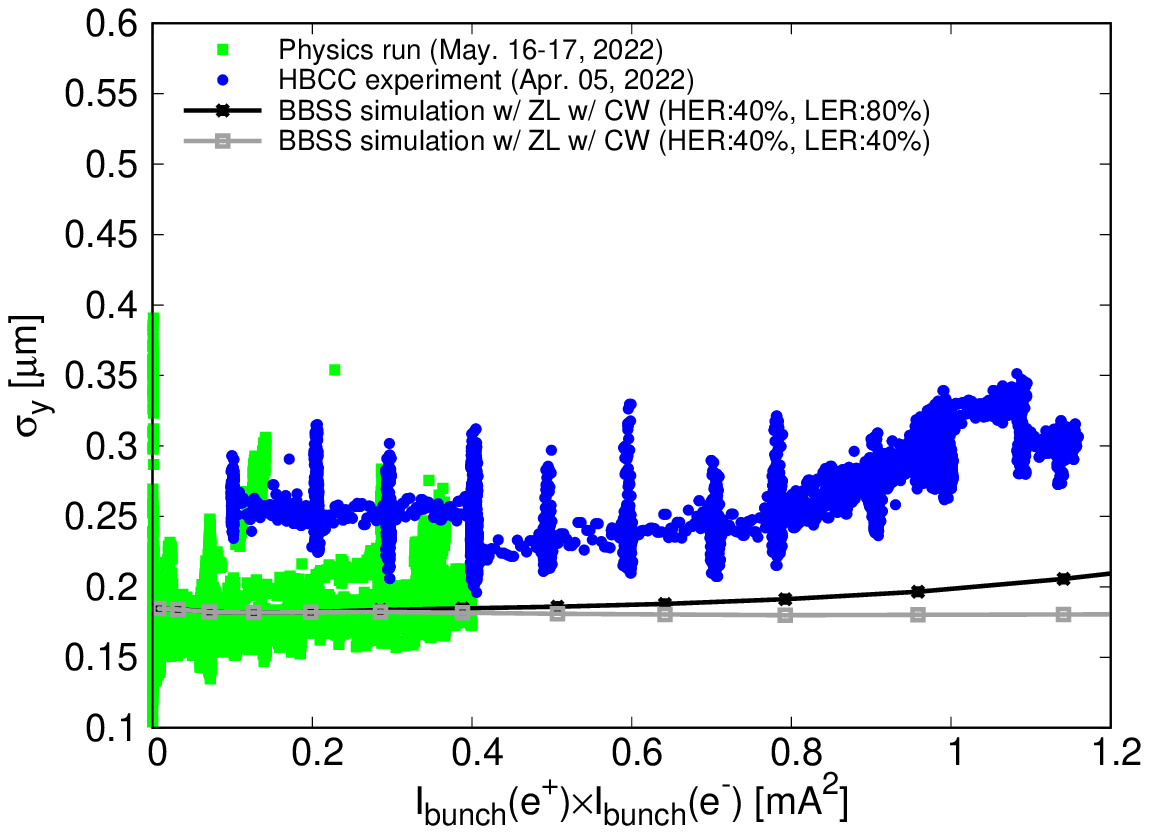}
    \vspace{-0mm}
    \vspace{-0mm}
    \includegraphics*[width=70mm]{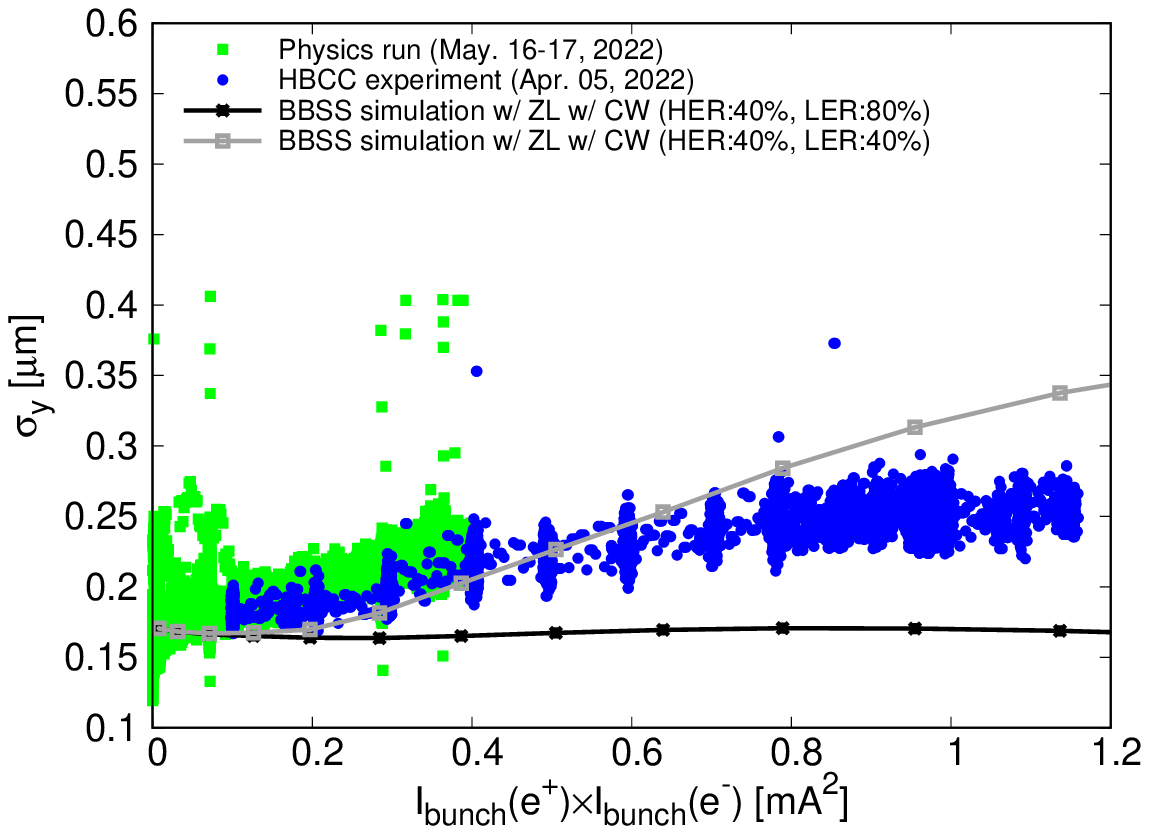}
    \vspace{0mm}
   \caption{Vertical beam sizes of the electron (upper) and positron (lower) beams at the IP, corresponding to Fig.~\ref{fig:Lum_Sim_vs_Exp_20220405}.}
   \label{fig:sigy_Sim_vs_Exp_20220405}
\end{figure}

\begin{figure}[!htb]
   \centering
   \vspace{-5mm}
    \includegraphics*[width=70mm]{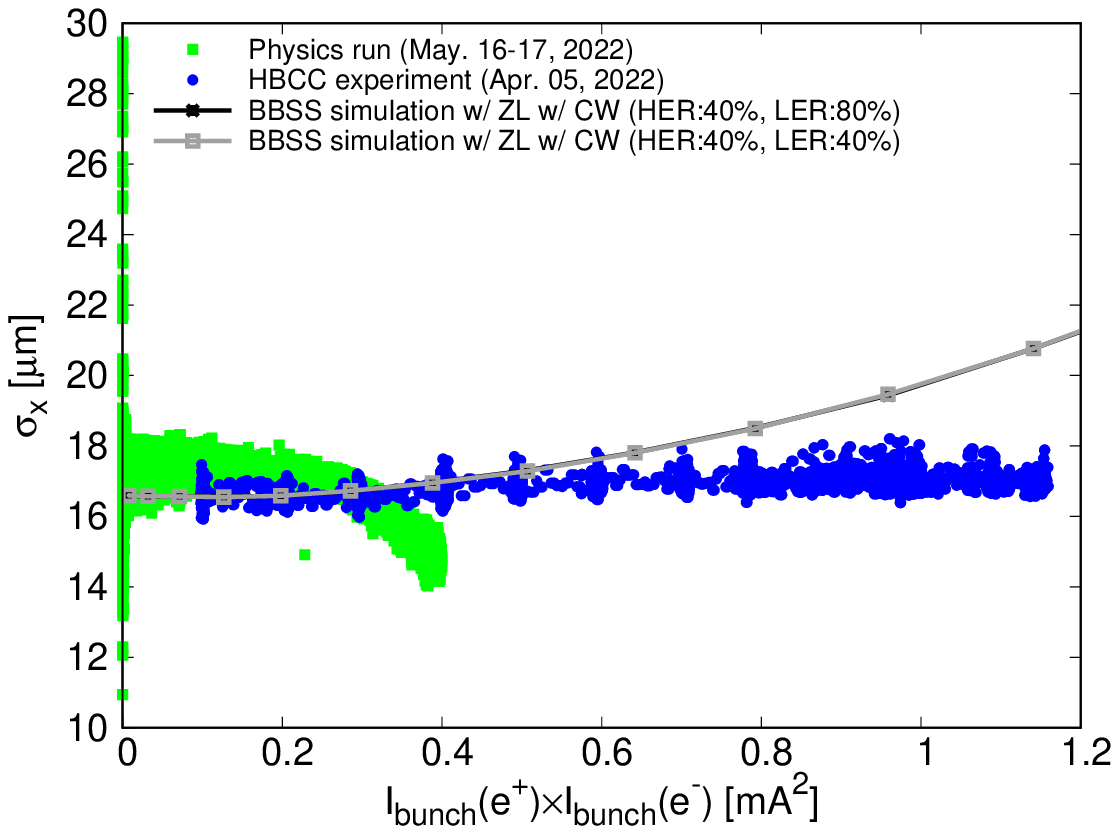}
    \vspace{-0mm}
    \vspace{-0mm}
    \includegraphics*[width=70mm]{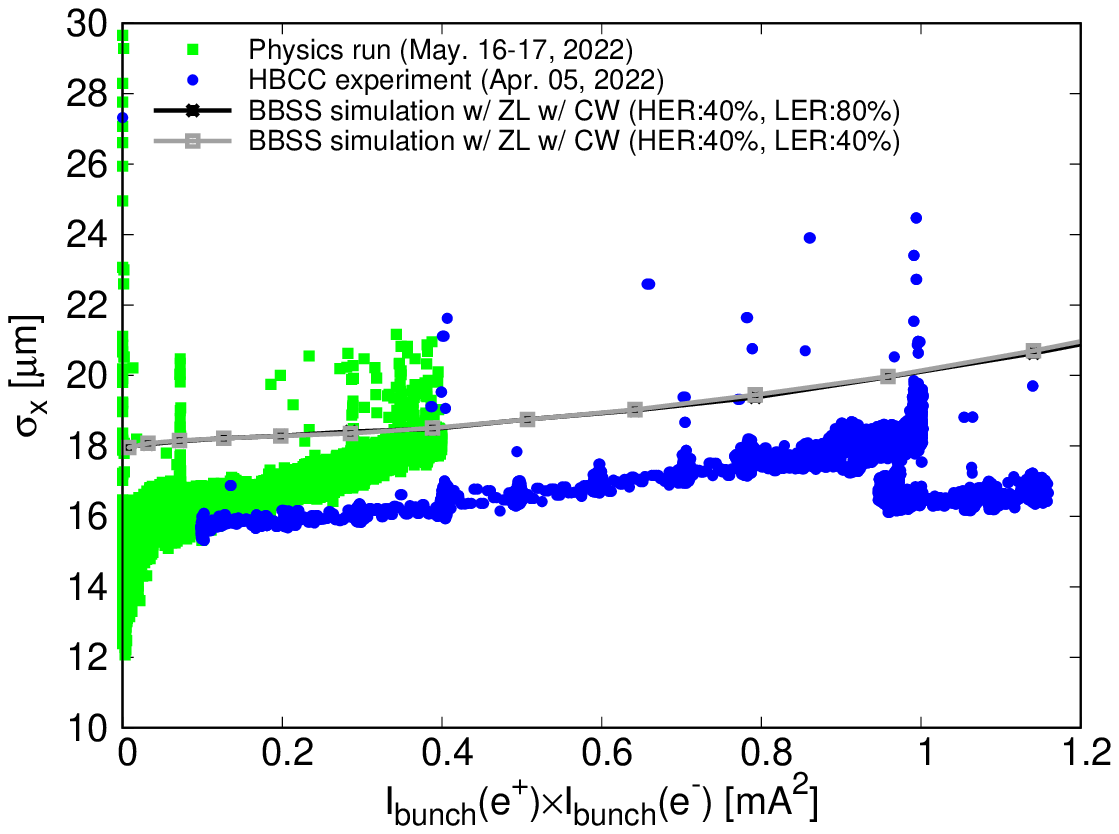}
    \vspace{0mm}
   \caption{Horizontal beam sizes of the electron (upper) and positron (lower) beams at the IP, corresponding to Fig.~\ref{fig:Lum_Sim_vs_Exp_20220405}.}
   \label{fig:sigx_Sim_vs_Exp_20220405}
\end{figure}

\begin{figure}[htb]
   \centering
    \vspace{-2mm}
   \includegraphics*[width=70mm]{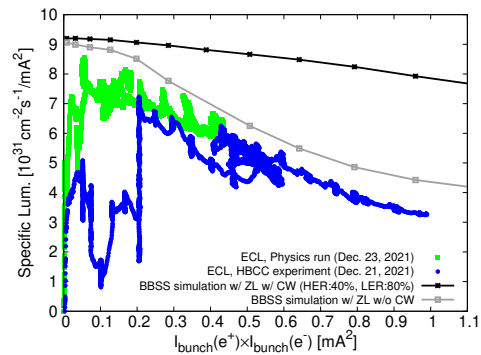}
    \vspace{0mm}
   \caption{Specific luminosity from HBCC machine study (blue dots) and physics run (green dots) measured by the ECL monitor in December 2021, compared to predictions of BBSS simulations with the inclusion of longitudinal impedances.}
   \label{fig:Lum_Sim_vs_Exp_20211221}
\end{figure}

Since April 2020, the crab waist has been implemented at SuperKEKB to suppress beam-beam resonances~\cite{Raimondi2007arXiv,Zobov2007PRL}. Luminosity performance has been improving with the following observations (see Refs.\cite{Ohnishi2021EPJP, Funakoshi2022IPAC, Ohnishi_eeFACT2022} for reviews): 1) Luminosity performance became closer to the predictions of simulations; 2) Balanced collisions (i.e., $\sigma_{y+}^* \approx \sigma_{y-}^*$) were achieved with careful tuning knobs; 3) The fractional working point could be set around the design values (.53, .57) (See Tab.~\ref{tb:parameters}.); 4) The total beam currents were not limited by beam-beam blowup, but by injection power and by machine failures such as sudden beam losses (SBLs, see Ref.~\cite{Ikeda_eeFACT2022} for details.); 5) There still exists an unexpected degradation of specific luminosity vs. product of bunch currents (see Figs.~\ref{fig:Lum_Sim_vs_Exp_20220405} and~\ref{fig:Lum_Sim_vs_Exp_20211221}). In particular, increasing the beam current does not give large increases in luminosity.

During the physics runs until June 2022, SuperKEKB has been operated with bunch currents less than about 0.7 mA and 0.57 mA (corresponding to a bunch-current product of about 0.4 mA$^2$) for the positron and electron beams, respectively. The limit on the bunch currents during the physics runs was mainly from the risks of machine failures due to SBLs. In 2021 and 2022, dedicated machine studies with 393 bunches (so-called high-bunch current collision (HBCC) machine studies with the number of bunches for collision much smaller than the usual physics run.) were done to extract the luminosity performance at higher bunch currents. When switching from the physics run to the HBCC machine study, we found that extra machine tunings (such as scans of closed orbit at the IP, scans of linear couplings at the IP, etc.) were necessary to optimize the luminosity. The main reason for such extra machine tunings was due to the current-dependent deformation of linear optics due to closed-orbit distortion caused by the synchrotron radiation heating (see Ref.~\cite{Ohnishi_eeFACT2022} for further details.).

\begin{figure}[!htb]
   \centering
   \vspace{-5mm}
    \includegraphics*[width=70mm]{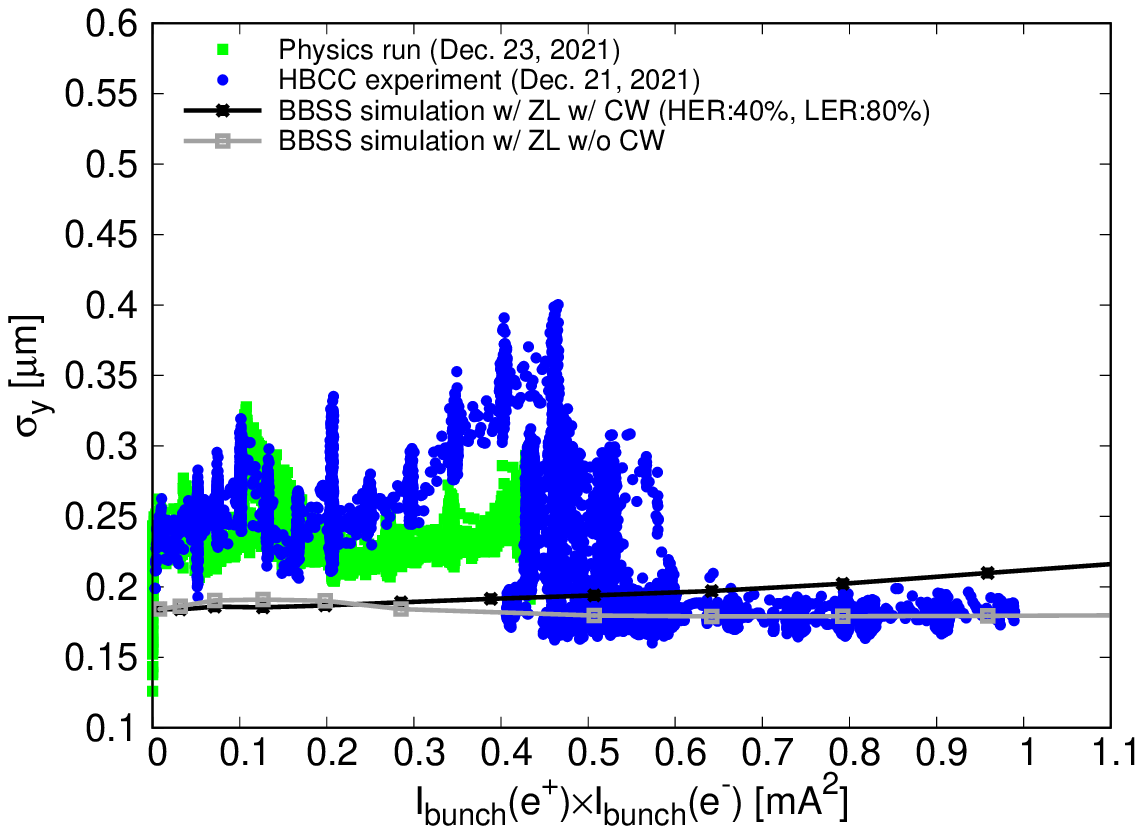}
    \vspace{-0mm}
    \vspace{-0mm}
    \includegraphics*[width=70mm]{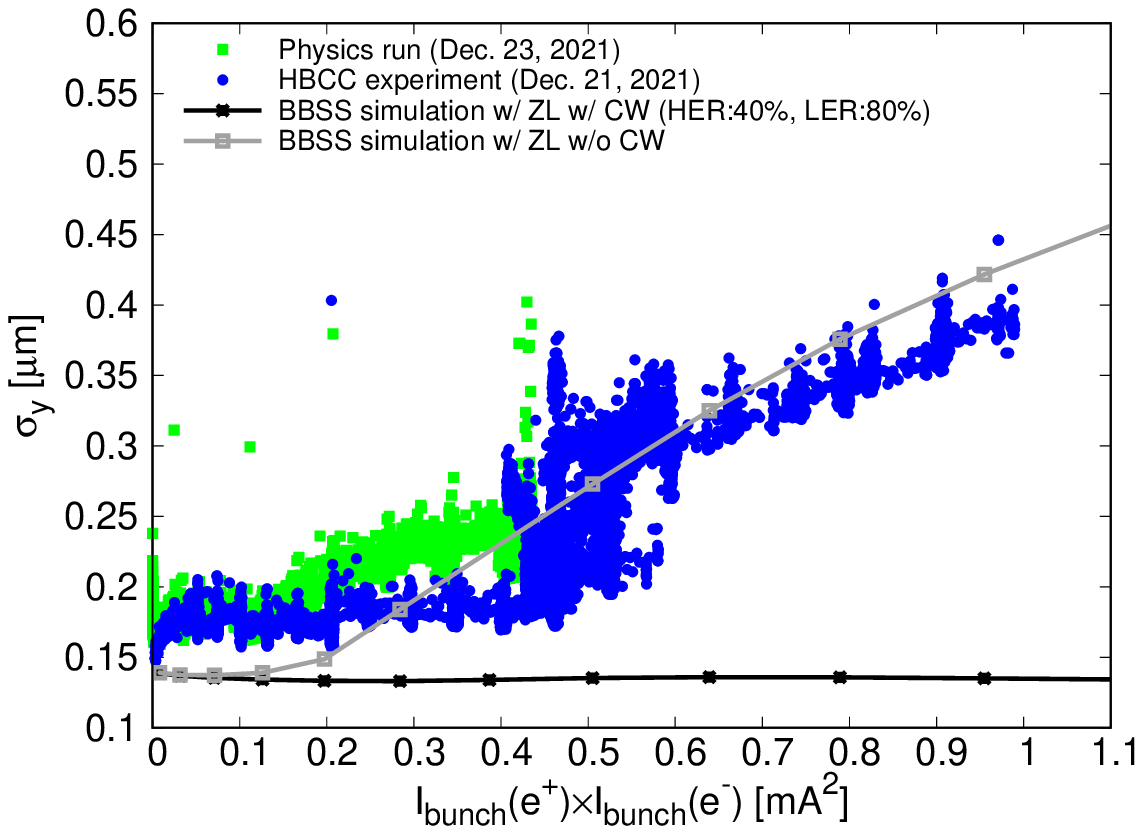}
    \vspace{0mm}
   \caption{Vertical beam sizes of the electron (upper) and positron (lower) beams at the IP, corresponding to Fig.~\ref{fig:Lum_Sim_vs_Exp_20211221}.}
   \label{fig:sigy_Sim_vs_Exp_20211221}
\end{figure}

\begin{figure}[!htb]
   \centering
   \vspace{-5mm}
    \includegraphics*[width=70mm]{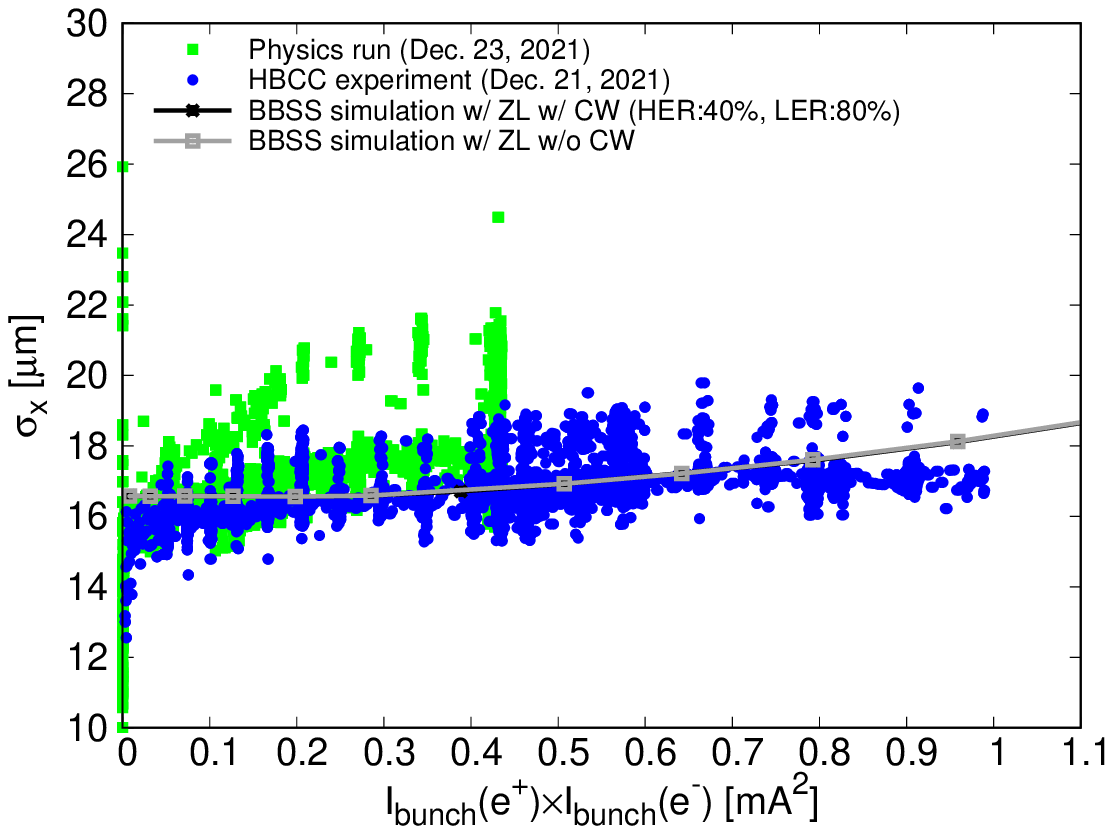}
    \vspace{-0mm}
    \vspace{-0mm}
    \includegraphics*[width=70mm]{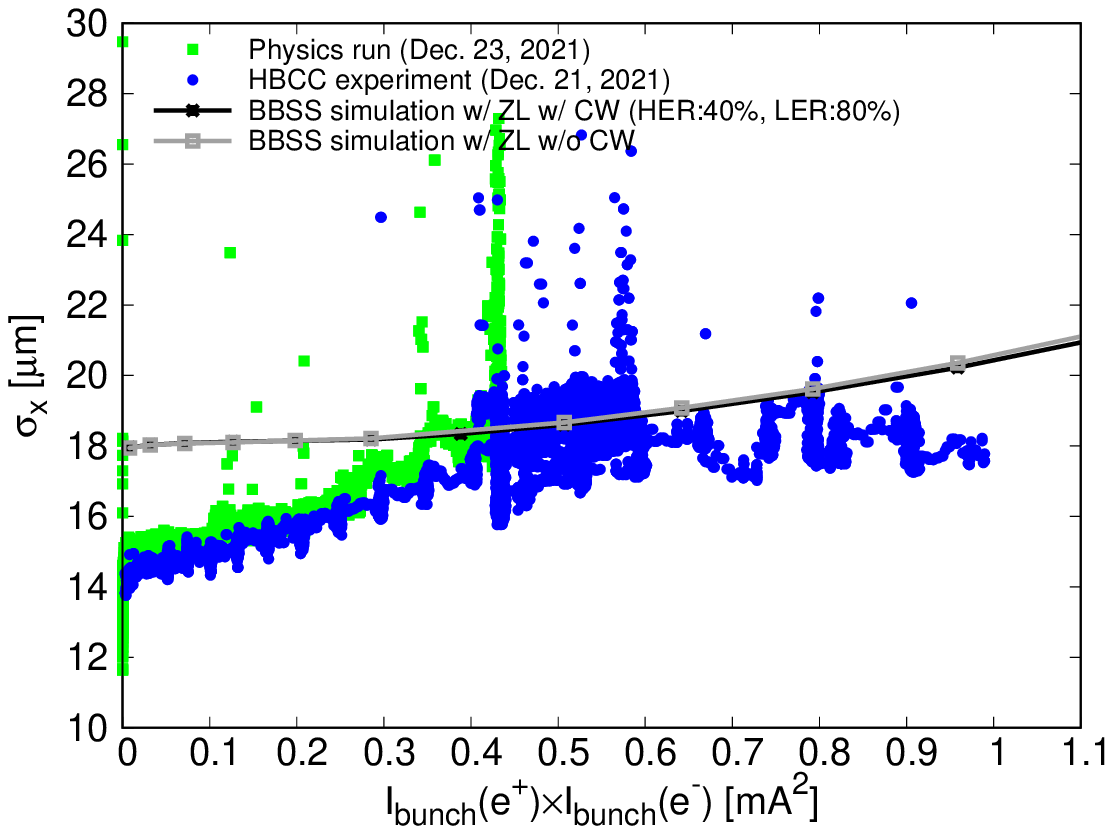}
    \vspace{0mm}
   \caption{Horizontal beam sizes of the electron (upper) and positron (lower) beams at the IP, corresponding to Fig.~\ref{fig:Lum_Sim_vs_Exp_20211221}.}
   \label{fig:sigx_Sim_vs_Exp_20211221}
\end{figure}

 Figures~\ref{fig:Lum_Sim_vs_Exp_20220405}-\ref{fig:sigx_Sim_vs_Exp_20211221} compare the specific luminosity and transverse beam sizes at the IP obtained from experiments (i.e., HBCC machine studies and physics runs) and from strong-strong beam-beam simulations.

 For the experiments under comparison, the global machine parameters were close to the parameter set of 2022.04.05, as shown in Tab.~\ref{tb:parameters}. Crab waist strengths of $R_{CW}=$ 40\% and 80\% were the standard settings for HER and LER in the experiments. The 40\% crab waist strength was set tentatively and can be increased in future commissioning. The electromagnetic calorimeter (ECL)~\cite{ECL2020CPC} has been used to measure the online luminosity at Belle II. The horizontal and vertical beam sizes for experiments were obtained using synchrotron radiation monitors (SRMs) and X-ray monitors (XRMs), respectively. Since the SRMs and XRMs are far from the IP, the beam sizes measured by those monitors do not represent the exact values at the IP. Indeed, the optics functions of relevant positions calculated from lattice models are used to estimate the beam sizes at the IP in experiments.

 For the simulations under comparison, the machine parameters refer to the set of 2022.04.05 in Tab.~\ref{tb:parameters}. The single-beam vertical emittances were $\epsilon_{y+}/\epsilon_{y-}$=20/35 pm for the simulations to compare the experiments as shown in Figs.~\ref{fig:Lum_Sim_vs_Exp_20211221}-\ref{fig:sigx_Sim_vs_Exp_20211221}. The longitudinal impedance causes bunch lengthening through potential-well distortion, and consequently reduces the luminosity according to Eq.~(\ref{eq:Lsp_Simple_approx1}). Therefore, the longitudinal impedances of both rings have been routinely included in beam-beam simulations.

Figures~\ref{fig:Lum_Sim_vs_Exp_20220405} and~\ref{fig:Lum_Sim_vs_Exp_20211221} show the specific luminosity measured in 2022 and 2021 with $\beta_y^*=1$ mm. In the HBCC machine studies, the collision for $I_{b+}I_{b-}\lesssim 0.4 \text{ mA}^2$ was not optimized due to limited beam time. Therefore, as a reference, we included the data from the physics runs of nearby dates, which represented the best performance in specific luminosity achieved with similar machine conditions as those of HBCC machine studies. In both simulations and experiments, the specific luminosity is sensitive to the vertical beam sizes at the IP, as described later. This is expected from the luminosity formulations in the previous section (for example, see Eq.~(\ref{eq:Lsp_Simple_approx1})). With crab waist strengths of $R_{CW}=$ 40\% and 80\% respectively for HER and LER, the decrease of specific-luminosity in strong-strong beam-beam simulation is mainly attributed to bunch lengthening due to the longitudinal wakefields and weak vertical blowup of the HER beam due to insufficient crab-waist strength. However, experimental results showed a much faster decrease as bunch currents increase. The plots also show simulations with the crab-waist strengths varied (see the gray lines of Fig.~\ref{fig:Lum_Sim_vs_Exp_20220405} for $R_{CW}=$40\% for both rings and Fig.~\ref{fig:Lum_Sim_vs_Exp_20211221} for $R_{CW}=0$ for both rings. In these simulations with reduced crab waist strengths, the fast drop of specific luminosity can be well understood: It is correlated with beam-beam-driven blowup in the positron beam (see the lower figure of Figs.~\ref{fig:sigy_Sim_vs_Exp_20220405} and \ref{fig:sigy_Sim_vs_Exp_20211221}), because its vertical fractional tune .589 is close to the 5th-order beam-beam resonances (as discussed in Sec.~\ref{sec:Lum_Performance_wo_CW}). Since the observed specific luminosity slope is closer to the simulations with crab waist strengths weaker than the values set to the rings, it tends to suggest that the crab waist settings might be imperfect in the machine operations. But, we have to point out that this is only one candidate to explain the specific luminosity slope observed in SuperKEKB. There have been other sources causing vertical blowup (and consequent luminosity degradation), but they are not included in the models of beam-beam simulations. Some of these sources will be discussed in the next section.

The optics setups were almost the same (i.e., $\beta_y^*$, working points, etc.) for the HBCC studies in 2021 and 2022, but the current-dependent vertical beam-size blowups were quite different, as shown in Figs.~\ref{fig:sigy_Sim_vs_Exp_20220405} and ~\ref{fig:sigy_Sim_vs_Exp_20211221}. One can see that the results of HBCC machine study in April 2022 showed gradual vertical beam-size blowup as the bunch currents were increased (see Fig.~\ref{fig:sigy_Sim_vs_Exp_20220405}); while in 2021, the vertical beam-size blowup was severe for both e+ and e- beams. At that time, it was difficult to achieve a balanced collision (i.e., $\sigma_{y+}^*\approx \sigma_{y-}^*$) through beam tunings: At the bunch-current products $I_{b+}I_{b-}\lesssim 0.4 \text{ mA}^2$, there was $\sigma_{y+}^* < \sigma_{y-}^*$; when $I_{b+}I_{b-}\gtrsim 0.4 \text{ mA}^2$, the positron beam blew up severely (see Fig.~\ref{fig:sigy_Sim_vs_Exp_20211221}). This swap of vertical blowup at high bunch currents was correlated with the ``-1 mode instability'' of the positron beam, which was driven by the interplay of vertical impedance (dominated by small-gap collimators) and the bunch-by-bunch (BxB) feedback (FB) system as discussed in detail in Ref.~\cite{Ohmi_eeFACT2022}. After fine-tuning the BxB FB system in March of 2022, the ``-1 mode instability'' was suppressed significantly, and the vertical beam-size blowup became less severe, as shown in Fig.~\ref{fig:sigy_Sim_vs_Exp_20220405}.

From both HBCC machine studies and beam-beam simulations, horizontal beam-size blowups have been observed (see Figs.~\ref{fig:sigx_Sim_vs_Exp_20220405} and~\ref{fig:sigx_Sim_vs_Exp_20211221}), though the specific luminosity is not sensitive to the horizontal beam sizes according to the luminosity formulations. The horizontal blowup in both beams observed in experiments had a qualitative agreement with beam-beam simulations. The decrease of the electron horizontal beam size shown in the upper subfigure of Fig.~\ref{fig:sigx_Sim_vs_Exp_20220405} was fake due to the failure in the XRM during that time. After fixing the XRM in HER, this phenomenon disappeared. The decrease of $\sigma_x^*$ for $I_{b+}I_{b-}>0.95 \text{ mA}^2$ as shown in the lower subfigure of Fig.~\ref{fig:sigx_Sim_vs_Exp_20220405} is due to the change of the horizontal tune of the e+ beam during the HBCC study. This $\nu_x$-dependence of horizontal beam-size blowup from the SRM data suggests that the weak horizontal blowup is driven by beam-beam interaction, not due to systematic error of the XRM monitor. Such horizontal-tune dependence of horizontal blowup was also seen in the HBCC study of Dec. 21, 2021 (see Fig.~\ref{fig:sigx_Sim_vs_Exp_20211221}): At bunch-current products of $I_{b+}I_{b-}>0.42 \text{ mA}^2$, the horizontal blowup of the e+ beam was remarkably relaxed. Meanwhile, improvement in the e+ beam's lifetime was observed, resulting in better injection efficiency. The mechanism of $\nu_x$-dependence of horizontal beam-size blowup can be explained as follows: As shown in Figs.~\ref{fig:TuneFootprintLER} and Tab.~\ref{tb:parameters}, after installing the crab-waist, both LER and HER have been operated with the horizontal tunes between the synchro-betatron resonances $\nu_x-\nu_s=N/2$ and $\nu_x-2\nu_s=N/2$. The beams' footprints spread in the tune space because of beam-beam, impedance effects, and lattice nonlinearity. When the tune footprint touches the resonance lines, the beam lifetime reduces, and extra beam losses appear in the injected bunches.


\begin{figure}[htb]
   \centering
    \vspace{-2mm}
   \includegraphics*[width=70mm]{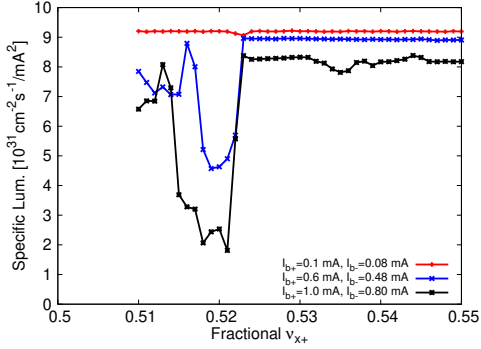}
    \vspace{0mm}
   \caption{Specific luminosity predicted by BBSS simulations with the inclusion of longitudinal impedances of both HER and LER and transverse impedance of only LER. Simulations were done by scanning the horizontal tune of LER and varying the bunch currents of the two beams. Other beam parameters are frozen the same as 2022.04.05 of Tab.~\ref{tb:parameters} except that $\epsilon_{y-}/\epsilon_{y+}=35/20\text{ pm}$ (Single-beam emittances observed on Dec. 21, 2021).}
   \label{fig:Lum_Sim_LER_nux_scan_20211221}
\end{figure}

\begin{figure}[!htb]
   \centering
   \vspace{-5mm}
    \includegraphics*[width=70mm]{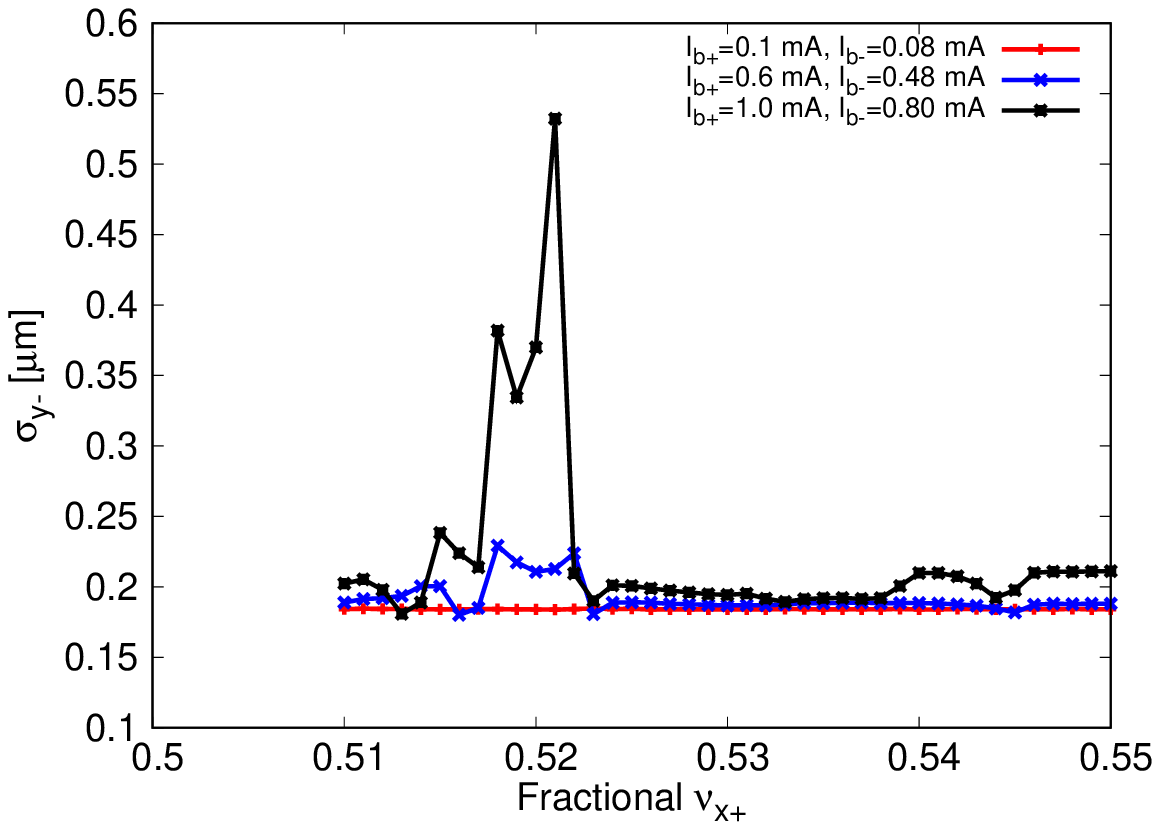}
    \vspace{-0mm}
    \vspace{-0mm}
    \includegraphics*[width=70mm]{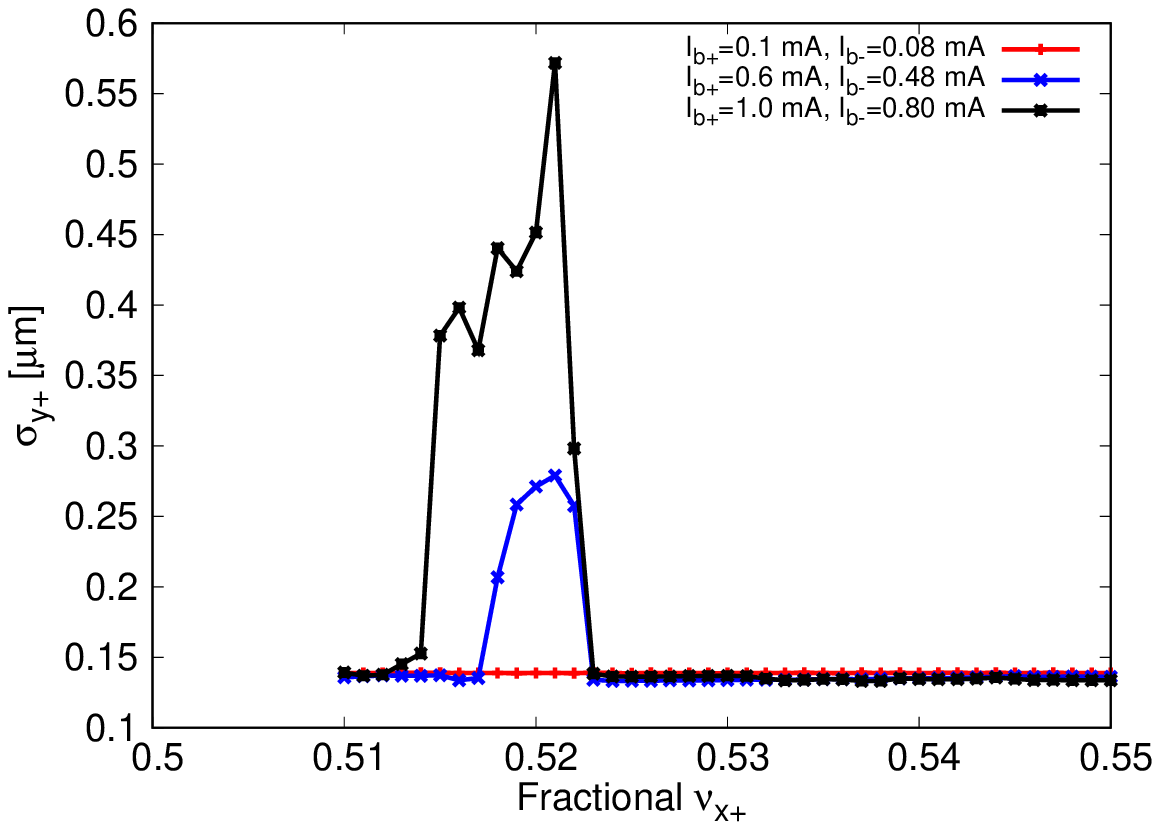}
    \vspace{0mm}
   \caption{Vertical beam sizes of electron (upper) and positron (lower) beams at the IP, corresponding to Fig.~\ref{fig:Lum_Sim_LER_nux_scan_20211221}.}
   \label{fig:sigy_Sim_LER_nux_scan_20211221}
\end{figure}

\begin{figure}[!htb]
   \centering
   \vspace{-5mm}
    \includegraphics*[width=70mm]{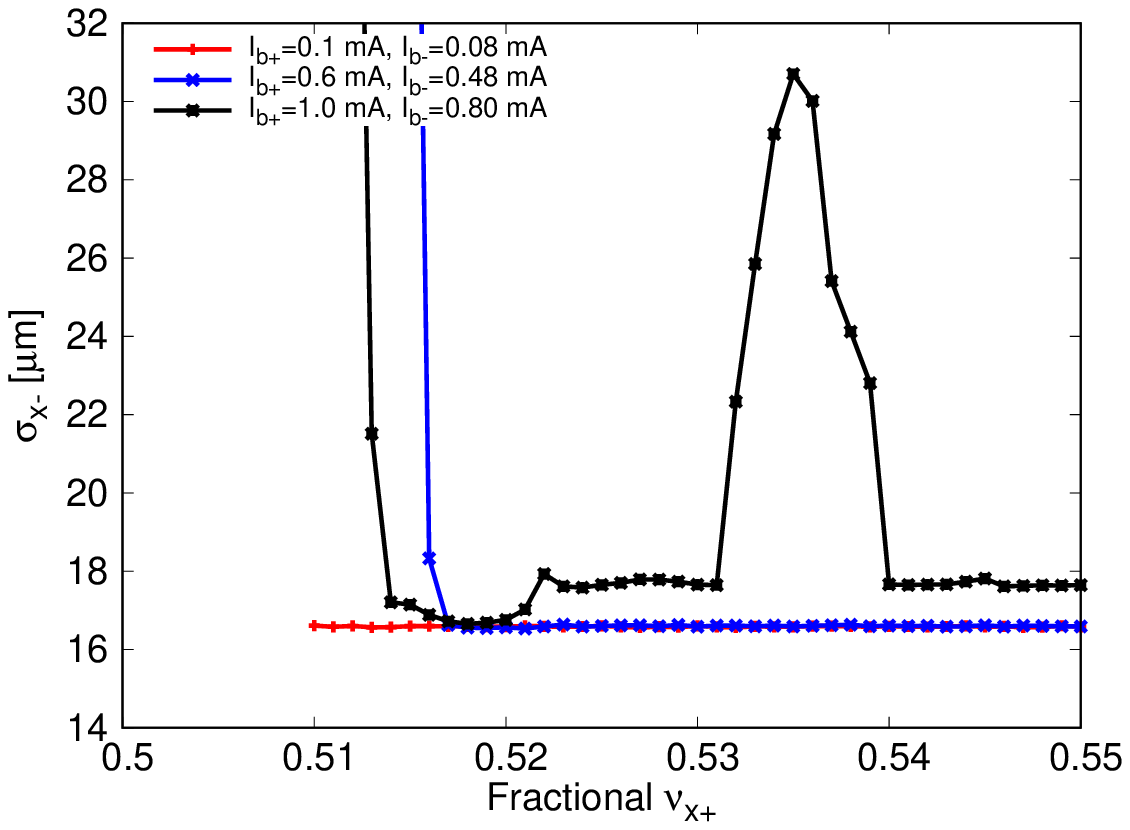}
    \vspace{-0mm}
    \vspace{-0mm}
    \includegraphics*[width=70mm]{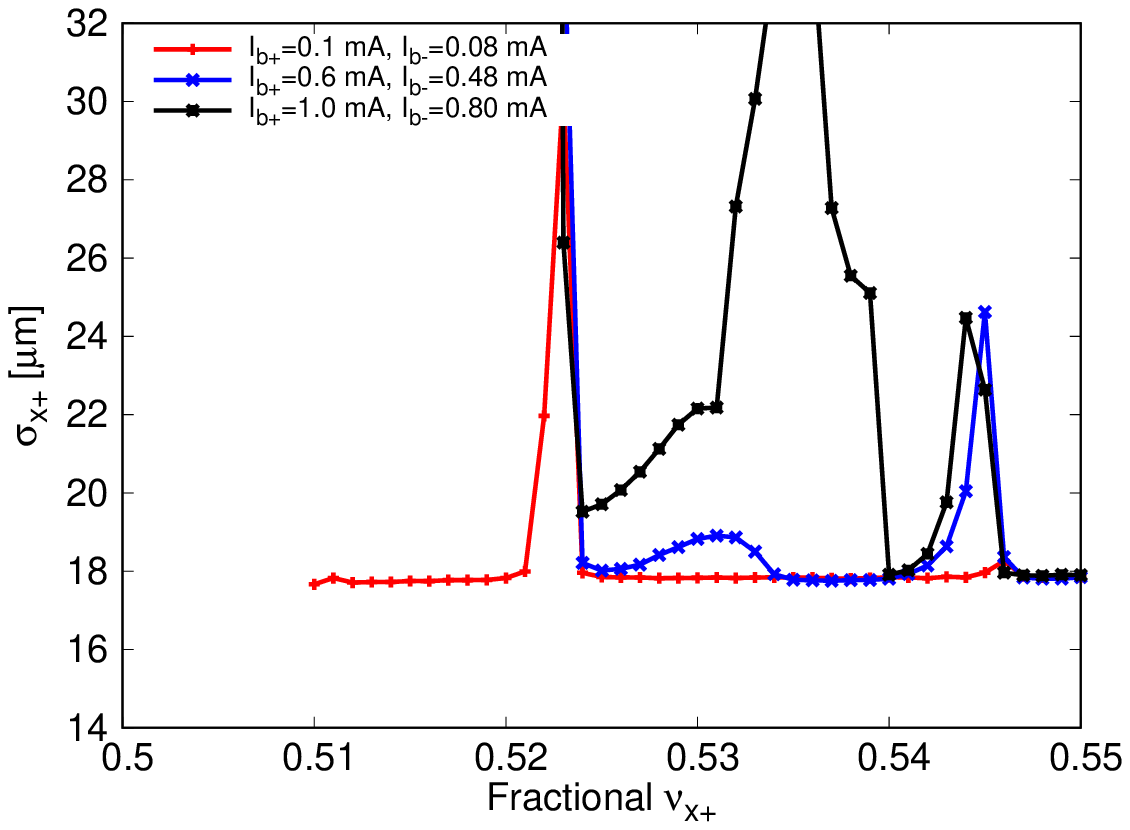}
    \vspace{0mm}
   \caption{Horizontal beam sizes of electron (upper) and positron (lower) beams at the IP, corresponding to Fig.~\ref{fig:Lum_Sim_LER_nux_scan_20211221}.}
   \label{fig:sigx_Sim_LER_nux_scan_20211221}
\end{figure}

Detailed strong-strong simulations have been done to investigate the tune-dependence of beam-beam effects on luminosity and beam sizes at SuperKEKB. In the following, we present simulation results using the BBSS code. Independent simulations were done using the IBB code~\cite{Zhang2022eeFACT2022}. Instability analysis was also performed, showing that the interplay of beam-beam interaction and the transverse impedance effects can drive a TMCI-like vertical instability~\cite{zhang_tmci_bb_2023}.

Figures ~\ref{fig:Lum_Sim_LER_nux_scan_20211221}, ~\ref{fig:sigy_Sim_LER_nux_scan_20211221} and ~\ref{fig:sigx_Sim_LER_nux_scan_20211221} show BBSS simulations (Scan of LER horizontal tune $\nu_{x+}$ with bunch currents varied) with the inclusion of longitudinal impedances of both HER and LER and transverse impedance of only LER (The transverse impedance of HER was not available when these simulations were done). The coherent BBHTI~\cite{Ohmi2017PRL, Kuroo2018PRAB} appears when the horizontal tune is close to the synchro-betatron resonances $2\nu_x-k\nu_s=N$. With the spread of incoherent synchrotron tune broadened by impedance effects~\cite{Lin2022PRAB}, the coherent BBHTI can appear in a large range of horizontal tunes~\cite{zhang2020self, migliorati2021interplay}. Even with the horizontal tune far from the synchro-betatron resonances, a weak and current-dependent blowup of horizontal beam size can appear, as shown in Fig.~\ref{fig:sigx_Sim_LER_nux_scan_20211221}. These simulations qualitatively agree with the experimental observations of horizontal blowups with collisions, as discussed previously.

Figure~\ref{fig:Lum_Sim_LER_nux_scan_20211221} shows that there is almost no change in the specific luminosity in the range $0.53\leq \nu_{x+}\leq 0.54$ while Fig.~\ref{fig:sigx_Sim_LER_nux_scan_20211221} shows large changes in the horizontal sizes over this range. This can be illustrated as follows. In Sec.~\ref{sec:Lum_formulations} we showed that when $\Phi_{HC}\gtrsim 1$ and  $\Phi_{XZ}\gg 1$ are satisfied, the specific luminosity can be well approximated by Eq.~(\ref{eq:Lsp_Simple_approx1}). The first condition can be explicitly written as
\begin{equation}
    \sigma_x^*\lesssim \frac{1}{\sqrt{2}} \beta_y^*\tan \frac{\theta_c}{2},
    \label{eq:lsp_condition1}
\end{equation}
and the second one is written as
\begin{equation}
    \sigma_x^*\ll \sigma_z \tan \frac{\theta_c}{2}.
    \label{eq:lsp_condition2}
\end{equation}
Here symmetric beams are assumed for simplification of the discussion. For Fig.~\ref{fig:Lum_Sim_LER_nux_scan_20211221}, $\beta_y^*=1$ mm and $\theta_c=83$ mrad, leading to $\sigma_x^*\lesssim 30$ $\mu$m. This is an estimate of the lower limit beyond which changes in $\sigma_{x\pm}^*$ affect the specific luminosity. One can see that the horizontal beam sizes shown in Fig.~\ref{fig:sigx_Sim_LER_nux_scan_20211221} coincide with this estimate. Since $\beta_y^*\ll \sigma_z$ is a nature of the nano-beam scheme, Eq.~(\ref{eq:lsp_condition2}) is automatically statisfied when Eq.~(\ref{eq:lsp_condition1}) is true.

\begin{figure}[htb]
   \centering
    \vspace{-2mm}
   \includegraphics*[width=70mm]{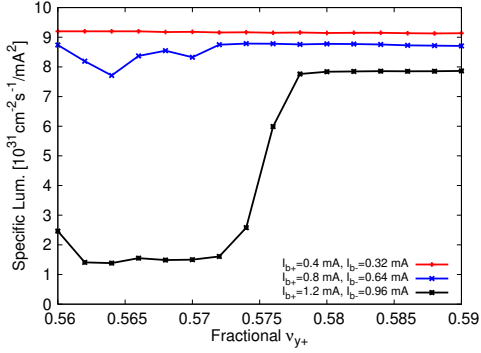}
    \vspace{0mm}
   \caption{Specific luminosity predicted by BBSS simulations with the inclusion of longitudinal impedances of both HER and LER and transverse impedance of only LER. Simulations were done by scanning the vertical tune of LER and varying the two beams' bunch currents. Other beam parameters are frozen the same as 2022.04.05 of Tab.~\ref{tb:parameters} except that $\epsilon_{y-}/\epsilon_{y+}=35/20\text{ pm}$ (Single-beam emittances observed on Dec. 21, 2021).}
   \label{fig:Lum_Sim_LER_nuy_scan_20211221}
\end{figure}

\begin{figure}[!htb]
   \centering
   \vspace{-5mm}
    \includegraphics*[width=70mm]{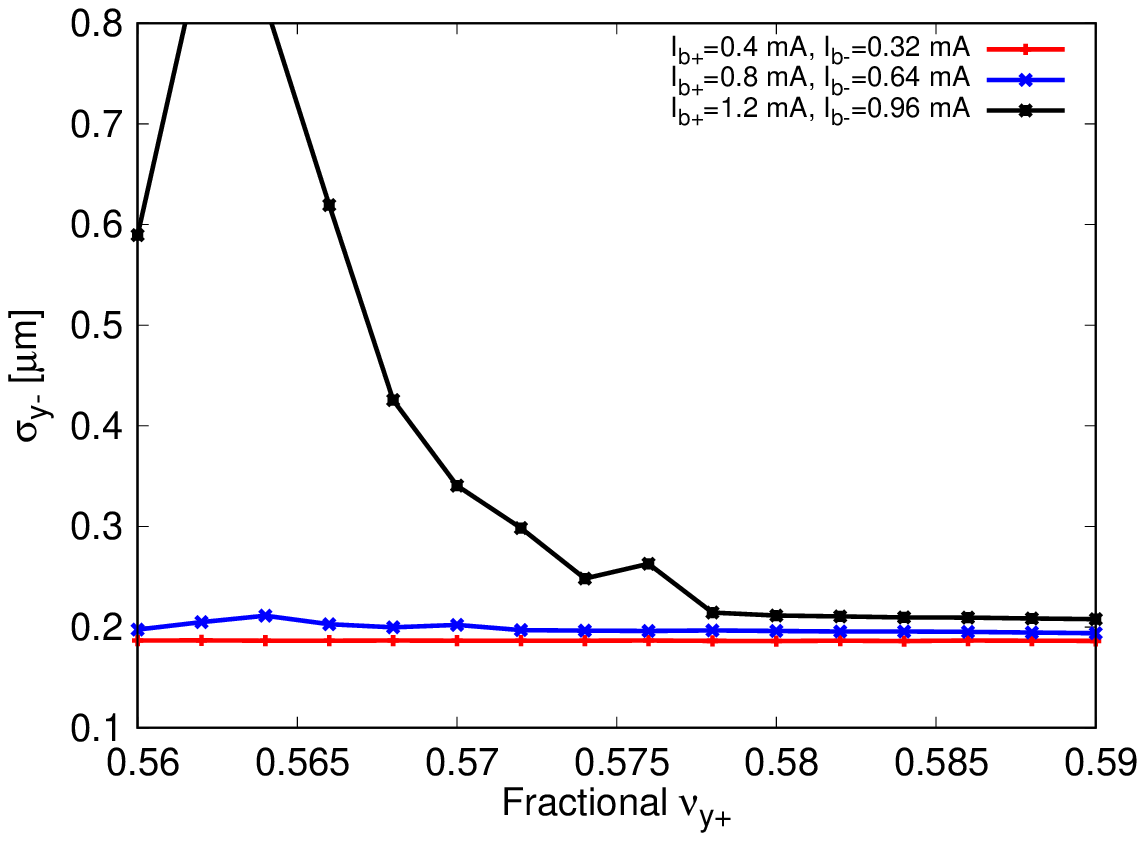}
    \vspace{-0mm}
    \vspace{-0mm}
    \includegraphics*[width=70mm]{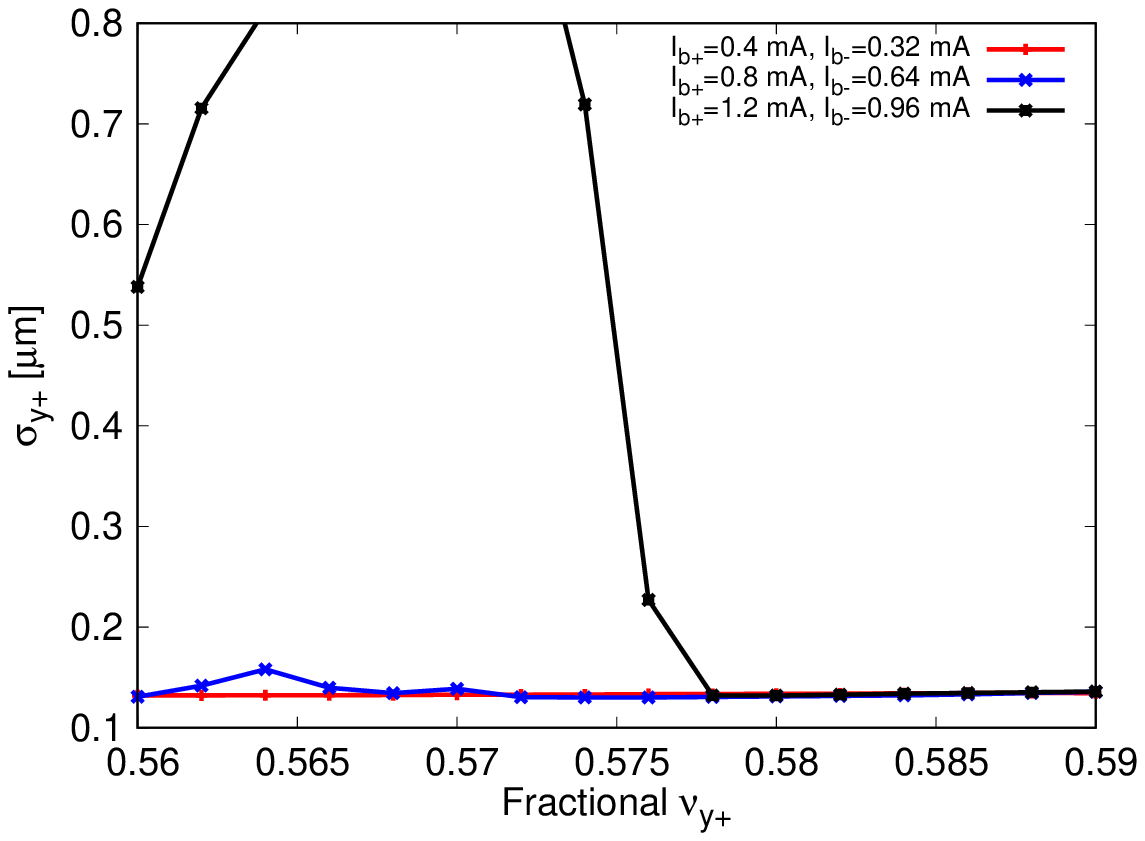}
    \vspace{0mm}
   \caption{Vertical beam sizes of electron (upper) and positron (lower) beams at the IP, corresponding to Fig.~\ref{fig:Lum_Sim_LER_nuy_scan_20211221}.}
   \label{fig:sigy_Sim_LER_nuy_scan_20211221}
\end{figure}

Figures ~\ref{fig:Lum_Sim_LER_nuy_scan_20211221} and~\ref{fig:sigy_Sim_LER_nuy_scan_20211221} show BBSS simulations by scanning the LER vertical tune $\nu_{y+}$ with bunch currents varied but the horizontal tunes frozen. The simulation conditions are the same as those for Fig.~\ref{fig:Lum_Sim_LER_nux_scan_20211221}. With the choices of horizontal tunes shown in Tab.~\ref{tb:parameters}, the coherent BBHTI does not appear in these simulations. But there are weak horizontal beam-size blowups which are current-dependent as seen in Fig.~\ref{fig:sigx_Sim_LER_nux_scan_20211221}. On the other hand, a vertical blowup can appear when the vertical tune approaches half-integer. This vertical blowup has a threshold current that is $\nu_y$-dependent. This vertical blowup driven by the interplay of beam-beam and vertical impedance was first discovered by K. Ohmi and Y. Zhang via simulations and later confirmed via instability analysis~\cite{zhang_tmci_bb_2023}. It is a $\sigma$-mode instability where the vertical betatron motion is perturbed by both the beam-beam force and the vertical conventional wake fields. The tune of the 0 mode is reduced by the ring impedance as the bunch current increases, while the tune of the -1 mode is increased by the cross wake of the beam-beam force. At a certain bunch current threshold, the 0 and -1 modes merge, and the TMCI-like $\sigma$-mode instability appears. The detailed analysis of this vertical instability is beyond the scope of this paper, and the reader is referred to Ref.~\cite{zhang_tmci_bb_2023}. The simulations shown here qualitatively agree with the experimental observation that the vertical fractional tune of LER could not approach the design value of 0.57 during machine operation. Meanwhile, beam-beam effects and lattice resonances (see Fig.~\ref{fig:TuneFootprintLER} as an illustration) require that the fractional vertical tune cannot be higher than 0.6.

\section{\label{sec:Source_of_Lum_Degradation}Sources of luminosity degradation}

\subsection{Known sources}
\begin{figure}[htb]
   \centering
    \vspace{-2mm}
   \includegraphics*[width=80mm]{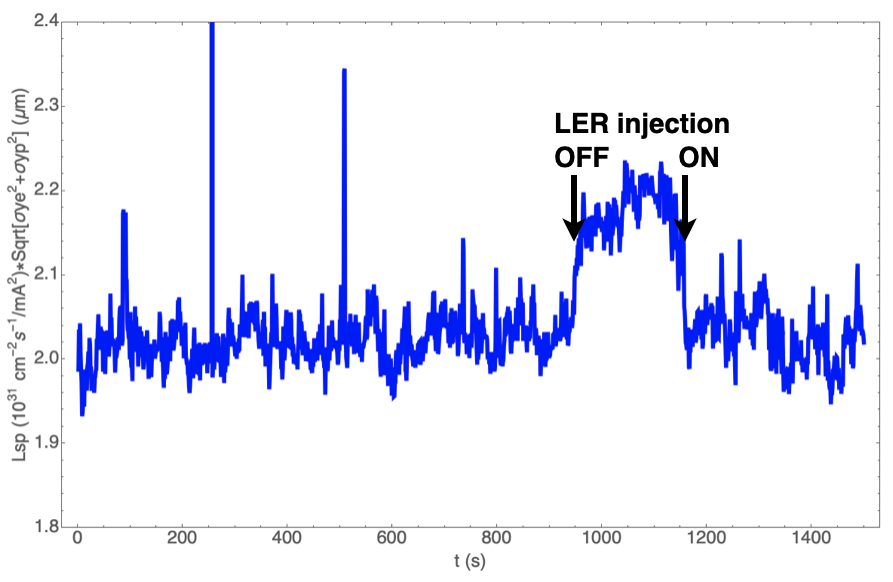}
    \vspace{0mm}
   \caption{The weighted luminosity $L_{sp}\Sigma_y^*$ synchronized with LER injection in SuperKEKB during the physics run on Jun. 2, 2022.}
   \label{fig:Lsp_drop_20220602}
\end{figure}

\begin{figure}[htb]
   \centering
   \vspace{-5mm}
    \includegraphics*[width=70mm]{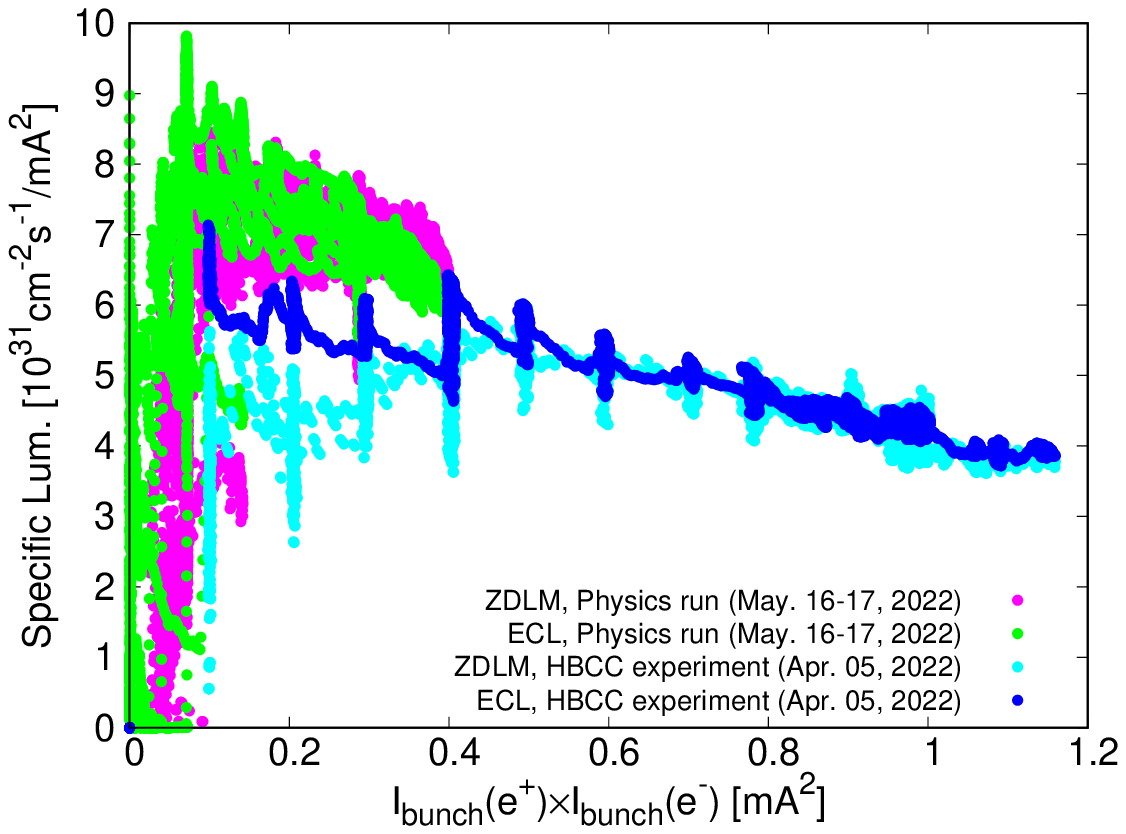}
    \vspace{0mm}
    \vspace{0mm}
    \includegraphics*[width=70mm]{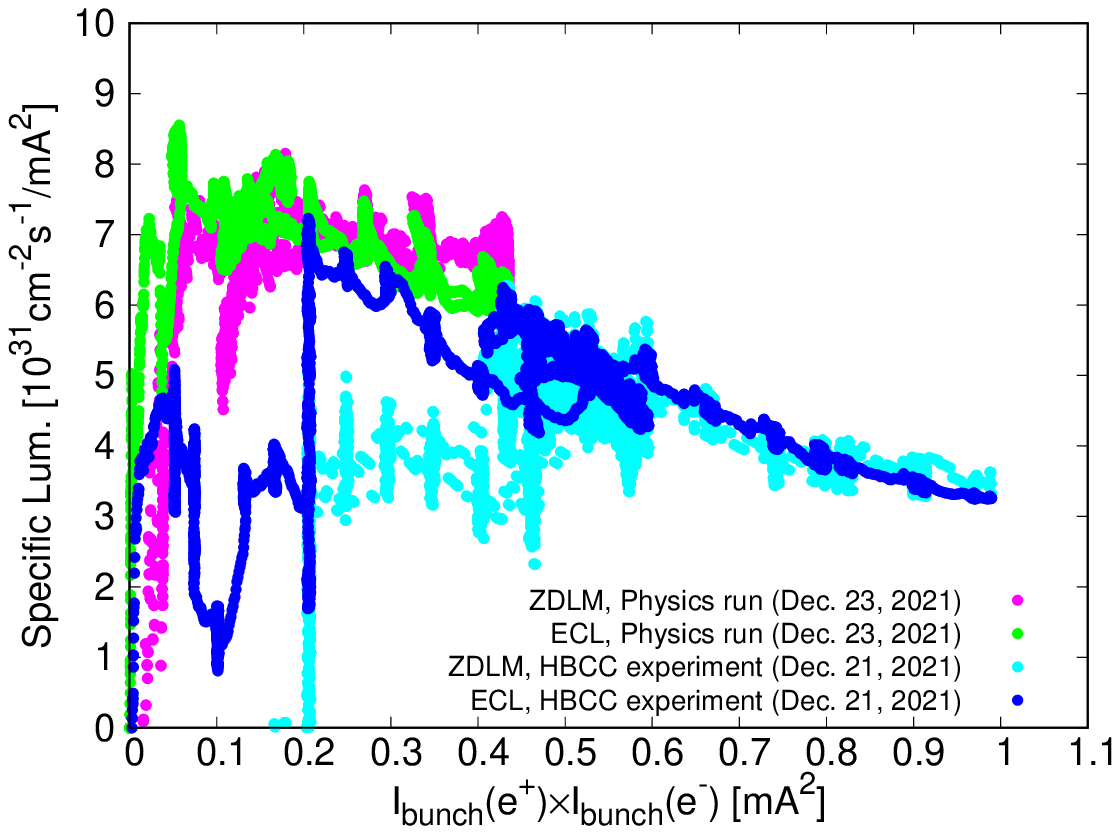}
    \vspace{0mm}
   \caption{Specific luminosity measured by ECL and ZDLM luminosity monitors, compared to predictions of BBSS simulations with the inclusion of longitudinal impedances. The upper and lower subfigures correspond to Figs.~\ref{fig:Lum_Sim_vs_Exp_20220405} and~\ref{fig:Lum_Sim_vs_Exp_20211221}, respectively.}
   \label{fig:Lum_Sim_vs_Exp_with_ZDLM_scaled}
\end{figure}

Simulations and experiments have identified some sources of luminosity degradation at SuperKEKB. The sources listed here are tentatively ordered from the most to the least important.
\begin{itemize}
    \item Bunch lengthening driven by longitudinal impedance. From Eq.~(\ref{eq:Lsp}), the specific luminosity follows the scaling law of $L_{sp} \propto 1/\Sigma_z$. Simulations using numerically constructed impedance models predict $\sigma_z(I_{b})=\sigma_{z0}+A\cdot I_{b}$ with $I_b$ the bunch current and $A$ about 1 mm/mA for both rings, while measurements using streak cameras showed $A$ to be about 2 mm/mA. The sources of discrepancy in simulated and measured bunch lengthening are under investigation. Nevertheless, the bunch lengthening is expected to cause a loss of geometric luminosity by order of 10\% at the bunch current product of $I_{b+}I_{b-}=1 \text{ mA}^2$.
    \item Vertical blowup in the LER driven by the interplay of vertical impedance and feedback system. The problem was well suppressed by fine-tuning the feedback system (see Refs.~\cite{Ohmi_eeFACT2022, Terui2022IPAC}). But this interplay can remain a source of vertical blowup, especially when the vertical small-gap collimators were severely damaged~\cite{Ishibashi_eeFACT2022}, generating extra vertical impedances.
    \item Chromatic couplings. Their effects on luminosity were recognized at KEKB~\cite{Zhou2010PRSTAB}. For SuperKEKB, rotatable skew-sextupoles are installed in LER, and dedicated skew-sextupoles are installed in HER to control the global chromatic coupling (see Ref.~\cite{Masuzawa2022IPAC} for further details). Simulations showed that chromatic couplings from the nonlinear IR can cause a remarkably large loss of luminosity if they are not well suppressed in the case of $\beta_{y+}^*/\beta_{y-}^*=0.27/0.3$ mm (i.e., the final design configuration of SuperKEKB, see Ref.~\cite{Hirosawa:IPAC18-THPAK099} for further details). For the case of $\beta_y^*=1$ mm (This is the achieved $\beta_y^*$ in 2021 and 2022), simulations with measured chromatic couplings showed a few percent of luminosity loss.
    \item Injection background. The luminosity data provided by ECL (it measures the absolute luminosity) is the most important reference for machine tunings and online optimizations at SuperKEKB. In the previous section, the beam-beam simulations are compared only with the ECL luminosity. In 2022, it was identified that the ECL luminosity had a clear correlation with the LER injection~\cite{Zhou2022CommissioningMeeting}. With the top-up injection, the injection background affected the ECL luminosity: The background increased when the total beam currents were higher, and the loss rate of ECL luminosity became higher at higher beam currents. Figure~\ref{fig:Lsp_drop_20220602} shows an example of this correlation. When the LER injection was intentionally turned off or on during the physics run, a sudden change in specific luminosity was observed. Investigations showed that the luminosity measurement by ECL was affected by the injection background during the LER's beam injection~\cite{Matsuoka2022CommissioningMeeting}. During the physics run with high total beam currents, the luminosity measured by ECL could drop by less than 5\% during LER injection while luminosity measured by ZDLM (Zero Degree Luminosity Monitor~\cite{Hirai2001NIMA}) did not~\cite{Matsuoka2022CommissioningMeeting}. This can be seen by comparing the ECL and ZDLM data as shown in Fig.~\ref{fig:Lum_Sim_vs_Exp_with_ZDLM_scaled}. Since the ZDLM monitor measures the relative luminosity but not the absolute luminosity, calibration of the ZDLM data is necessary to compare the ECL data. In the comparison of Fig.~\ref{fig:Lum_Sim_vs_Exp_with_ZDLM_scaled}, we assumed a linear correlation between the ECL and ZDLM data and then scaled the ZDLM data to match the ECL data at low beam currents. After this scaling, we found that the ZDLM luminosity is higher than the ECL one during the physics runs with high beam currents (see the difference between magenta and green dots at $I_{b+}I_{b-}\gtrsim 0.3 \text{ mA}^2$ in Fig.~\ref{fig:Lum_Sim_vs_Exp_with_ZDLM_scaled}). On the other hand, the ECL and ZDLM data from the HBCC machine studies had a good agreement (see the blue and cyan dots at $I_{b+}I_{b-}\gtrsim 0.4 \text{ mA}^2$ in Fig.~\ref{fig:Lum_Sim_vs_Exp_with_ZDLM_scaled}). This is because the total beam current was lower than 500 mA, and the background to the Belle II was low enough and did not affect the ECL luminosity. In the end, we concluded that this luminosity-background correlation observed at ECL was fake and irrelevant to beam-beam interaction. Further investigations are ongoing to understand the correlation between ECL luminosity and the background from LER injection. This specific-luminosity-injection correlation through injection background indirectly impacts the beam dynamics by affecting online luminosity optimization via machine tunings. A calibration algorithm is planned to correct the online ECL luminosity data. In principle, this fake luminosity loss can be removed in future physics runs at SuperKEKB.
    \item Beam oscillation excited by the injection kickers of LER. It was found that the injection kickers in the LER were not perfectly balanced. This causes a leakage kick to the beam in the horizontal direction during the injection. Due to the global coupling of the lattice, the vertical oscillation is also excited. From the waveform of the kickers' field, roughly 20\% of the stored beam will be excited. The BxB FB system can damp the dipole oscillations in less than 200 turns (Compared with the radiation damping time of about 4500 turns). A simple estimate shows it will cause a loss rate of about 1\% to the luminosity.
\end{itemize}

\subsection{Sources to be investigated}
\begin{figure}[htb]
   \centering
    \vspace{-2mm}
   \includegraphics*[width=70mm]{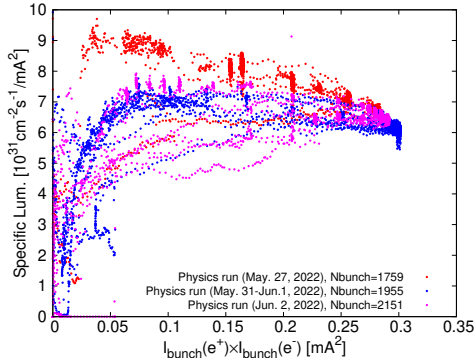}
    \vspace{0mm}
   \caption{Measured specific luminosity as a function of bunch current product with the different numbers of bunches during the physics runs in 2022. Machine tunings were routinely done to achieve the best luminosity performance around $I_{b+}I_{b-}\approx 0.3 \text{ mA}^2$.}
   \label{fig:Lum_Comparison_CBI_20220527}
\end{figure}

There are sources of luminosity degradation to be investigated through simulations and experiments:
\begin{itemize}
    \item Imperfect crab waist and insufficient crab-waist strengths. The nonlinear optics and optics distortion (its sources include machine errors, current-dependent orbit drift, etc.) around the IR might reduce the effectiveness of crab waist in suppressing beam-beam resonances. In 2022, it was identified that the synchrotron radiation (SR) heating caused drift of closed orbit (COD) at SuperKEKB~\cite{Ohnishi_eeFACT2022}. The small horizontal offset at the strong sextupoles for local chromaticity correction generates a significant beta-beat in the rings. Figures ~\ref{fig:Lum_Sim_vs_Exp_20220405} and~\ref{fig:Lum_Sim_vs_Exp_20211221} show luminosity degradation by insufficient crab-waist strengths. The crab-waist strength of HER has been set at 40\%.  BBWS simulations showed that this is insufficient to suppress the 5th-order beam-beam resonances and can be a source of vertical blowup in the electron beam and consequent luminosity degradation (See Fig.~\ref{fig:Lum_tune_scan_HER_CW_20210701} for a comparison of simulated luminosity with crab-waist strengths 0.4 and 0.8 in the HER.). Changing the crab waist strengths and consequent beam tunings must be done in future commissioning. In Particular, the dispersion functions in the IR need to be better controlled. Meanwhile, sextupole settings in the IR should be optimized considering both the local chromaticity correction and crab waist strengths.
\begin{figure}[!htb]
   \centering
   \vspace{-9mm}
    \includegraphics*[width=80mm]{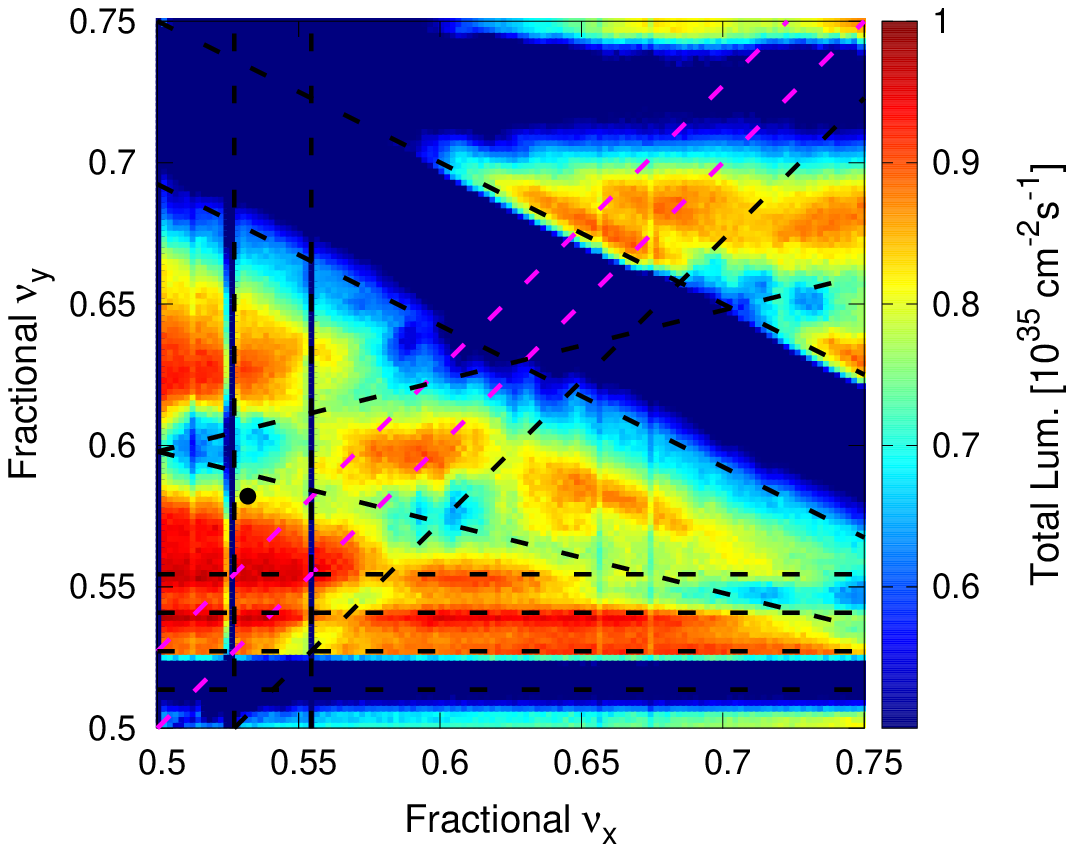}
    \vspace{0mm}
    \vspace{0mm}
    \includegraphics*[width=80mm]{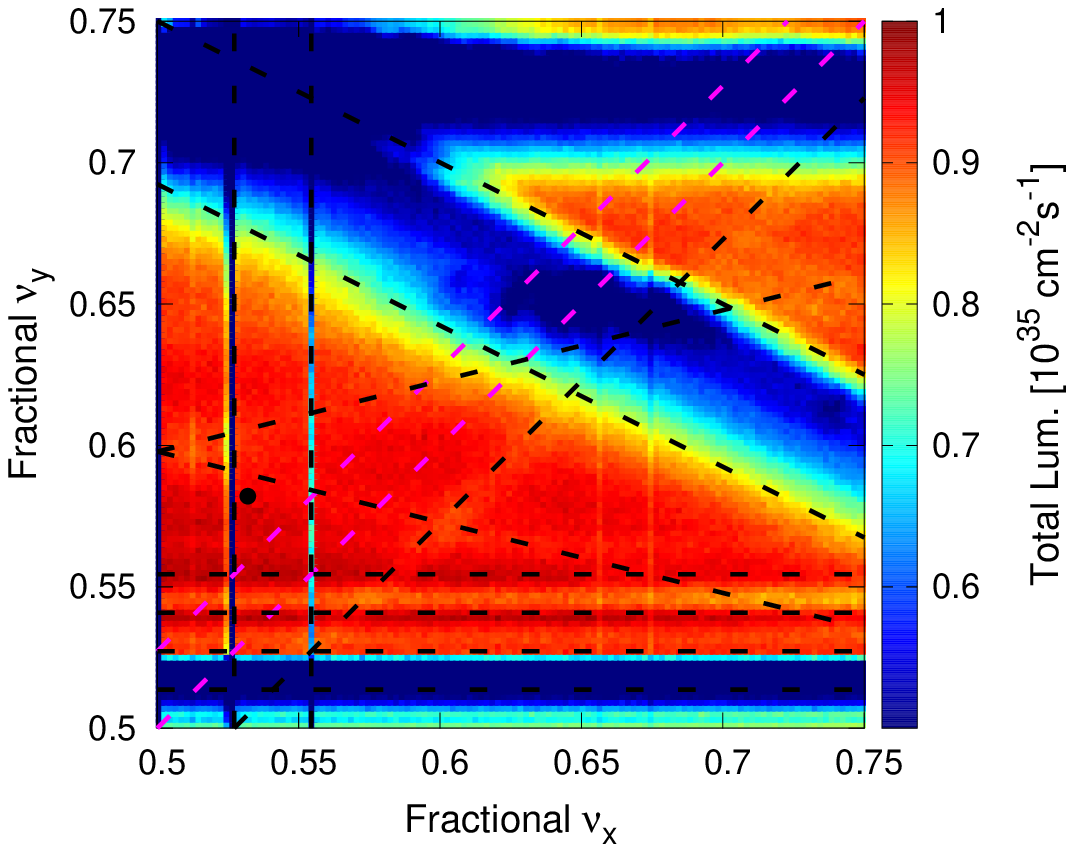}
    \vspace{-4mm}
   \caption{Tune scan of luminosity with crab-waist strength 0.4 (upper) and 0.8 (lower) for HER, $I_+/I-=1.0/0.8 \text{ mA}$, $\epsilon_{y+}/\epsilon_{y-}=23/23\text{ pm}$, and other beam parameters are referring to the parameter set of 2022.04.05 in Tab.~\ref{tb:parameters}. The HER beam is taken as the weak beam in the BBWS simulation. Important resonant lines are plotted, and the black dot indicates the working point for machine operation.}
   \label{fig:Lum_tune_scan_HER_CW_20210701}
\end{figure}
    \item BB-driven incoherent synchro-betatron resonances. Currently, the working point of SuperKEKB is between $\nu_x-\nu_s=N/2$ and $\nu_x-2\nu_s=N/2$ (See Figs.~\ref{fig:TuneFootprintLER} and Tabl~\ref{tb:parameters}), which are strong due to the beam-beam interaction~\cite{Raimondi2007arXiv} and nonlinear chromatic optics. The tune space in this region might not be large enough to hold the footprint of the beams. Note that collective effects and machine nonlinearity stretch the tune footprint.
    \item Interplay of beam-beam, longitudinal and transverse impedances, and BxB FB system. The interplay of transverse impedances and BxB FB system is discussed in Refs.~\cite{Terui2022IPAC,Ohmi_eeFACT2022}. To simulate the interplay of all these three factors, it is necessary to construct a realistic model of FB system, taking into account the realistic settings of the FB parameters, the environment noises, etc.
    \item Interplay of beam-beam and nonlinear lattices. This was identified as important for the final design of SuperKEKB configurations but should not be for the case of $\beta_y^*=$1 mm~\cite{Zhou2015IPAC}. On this issue, the machine errors are unknown sources of lattice nonlinearity. The crab waist, not counted in the final design, introduces additional nonlinearity to the lattices.
    \item Coupled bunch instabilities (CBI) with large bunch numbers and high total currents. With 2151 bunches and total beam currents of 1.4/1.12 A achieved in LER/HER, specific luminosity degradation due to CBI has not been seen. As shown in Fig.~\ref{fig:Lum_Comparison_CBI_20220527}, machine tunings with different numbers of bunches for collisions led to the same best luminosity. This indicates that CBI, which is always suppressed by the BxB FB system, should not be a source of specific-luminosity degradation in the current phase. Furthermore, the ZDLM luminosity data showed flat BxB luminosity~\cite{Uehara_private2022}, and CBI driven by electron cloud was not observed for the cases shown in Fig.~\ref{fig:Lum_Comparison_CBI_20220527}. Even with these observations, CBI at higher total currents (e.g., close to the design values 3.6/2.6 A in LER/HER) remains a concern. Therefore, we keep CBI on the list of sources to be investigated.
\end{itemize}

The sources listed above define the challenges and direction toward developing a predictable model of luminosity simulation.

\section{\label{sec:BB_Parameters}Beam-beam parameters}

For the convenience of discussion, we use $\xi_{y\pm}^L$ as formulated in Sec.~\ref{sec:Intro} to discuss the beam-beam parameters achieved with $\beta_y^*=1$ mm at SuperKEKB. Given beam sizes at the IP, the incoherent beam-beam tune shift $\xi_{y\pm}^{ih}$ can be calculated according to Eq.~(\ref{eq:BB_paramemter_with_Hourglass_effect1}).

As stated in Tab.~1 of Ref.~\cite{Funakoshi2022IPAC}, the achieved $\xi_{y\pm}^L$ during the physics run of SuperKEKB (i.e., the high voltage of Belle II was on.) in 2022 were 0.0407 and 0.0279 in LER and HER, respectively. During the physics run, the strategy of machine tunings for luminosity optimization was to achieve the best specific luminosity with $\sigma_{y+}^*\approx\sigma_{y-}^*$, but without the constraint of energy transparency condition $\gamma_+I_{b+}=\gamma_-I_{b-}$. This is the main reason for unequal beam-beam parameters observed at LER and HER.

Though the achieved beam-beam parameters were much lower than the design values, as shown in Tab.~\ref{tb:parameters:LumTest}, it does not mean that the beam-beam limit was already reached at SuperKEKB. During the physics run until June 2022, the main obstacles to storing high total beam currents for collisions were 1) the high risks of sudden beam losses and 2) the short beam lifetime and insufficient beam injection power~\cite{Funakoshi2022IPAC}.

From the HBCC machine studies with $\beta_y^*=1 \text{ mm}$, the highest beam-beam parameters achieved in 2022 were 0.0565 and 0.0434, respectively, for LER and HER with the BxB FB system off~\cite{Funakoshi2022IPAC}. As can be seen, the HBCC results are higher than that of physics runs, although they are still lower than the design values. Higher beam-beam parameters are expected to be achievable at SuperKEKB when the sources of luminosity degradation discussed in the previous section are better understood and effective remedies are developed.

\section{\label{sec:Summary}Summary and outlook}

Since April 2020, the crab waist has been incorporated with the nano-beam collision scheme at SuperKEKB and has proved decisive in suppressing nonlinear beam-beam effects. Future machine tunings and upgrades of SuperKEKB are expected to go with the crab waist. Though there is a strong interplay between beam-beam, crab waist, and lattice nonlinearity with the final design configuration~\cite{MoritaICFABD2015}, the crab waist should be tolerable with $\beta_y^*\geq 0.6 \text{ mm}$~\cite{Oide2021CWdesign} at SuperKEKB.

The interplay between beam-beam and single-bunch impedance effects is critical at SuperKEKB. Especially the longitudinal monopole and vertical dipole impedances are essential in affecting machine performance. The intense interplay of bunch-by-bunch feedback and vertical impedance in LER has been a strong limit of luminosity performance until April 2022. After fine-tuning the feedback system, this problem was relaxed but remained a possible source of mild vertical emittance blowup.

With progress in machine tuning, some sources of luminosity degradation with the crab waist have been well-identified. The measured luminosity and beam sizes with $\beta_y^*=1 \text{ mm}$ have been approaching the predictions of beam-beam simulations. However, it is also true that the existing simulation tools cannot fully predict the machine parameters. Including multiple beam dynamics (such as beam-beam, crab waist, impedances, lattice imperfections, and BxB FB) in the beam-beam simulations is required, especially to predict the luminosity at high beam currents and smaller $\beta_y^*$.

In addition to the factors discussed in this paper, the space-charge (SC) effects in the LER can be important in affecting the luminosity performance of SuperKEKB. When the positron beam has low emittance and high bunch currents, the SC-driven tune shift can be comparable with beam-beam tune shifts. Simulations with a weak-strong model of SC were done and showed a significant luminosity degradation with the baseline configurations of SuperKEKB~\cite{Zhou2015IPAC, ZhouICFABD2015}. Conclusive simulations require a self-consistent model of SC implemented in the strong-strong beam-beam simulations with the full lattices.

\section{ACKNOWLEDGEMENTS}
We thank the SuperKEKB and the Belle II teams for their constant support of our work. Since July 2021, an international task force (ITF) has been organized to work on beam physics at SuperKEKB. We thank the ITF members (especially K. Oide (CERN), D. Shatilov (BINP), M. Zobov (INFN), T. Nakamura (J-PARC), T. Browder (UH), Y. Cai (SLAC), C. Lin (IHEP), et al.) for their contributions. The author D.Z. thanks S. Uehara and K. Matsuoka for fruitful discussions on luminosity measurements at Belle II.


\bibliography{aps_bb_skb}

\end{document}